\documentclass[onecolumn,showpacs,amsfonts,aps,prc,nofootinbib,floatfix,%
superscriptaddress]{revtex4}

\usepackage{amsmath}
\usepackage{bm}
\usepackage{graphicx}

\usepackage{epsfig}
\newcommand{\beq}{\begin{equation}}
\newcommand{\eeq}{\end{equation}}
\newcommand{\bea}{\vspace{0.25cm}\begin{eqnarray}}
\newcommand{\eea}{\end{eqnarray}}

\newcommand{\ro}{\mbox{{\boldmath
$\rho$}}}

\newcommand{\qb}{\mbox{{\bf
q}}}

\newcommand{\bb}{{{\bf b}}}

\newcommand{\qbt}{\mbox{{\bf
q}}_\perp}

\def\lsim{\mathrel{\rlap{\lower4pt\hbox{\hskip1pt$\sim$}}
    \raise1pt\hbox{$<$}}}         
\def\gsim{\mathrel{\rlap{\lower4pt\hbox{\hskip1pt$\sim$}}
    \raise1pt\hbox{$>$}}}         

\newcommand{\landau}{L.D.~Landau Institute for Theoretical Physics,
        GSP-1, 117940, Kosygina Str. 2, 117334 Moscow, Russia}

\begin{document}


\title{
Color randomization of fast gluon-gluon pairs in the quark-gluon plasma
}
\date{\today}

\author{B.G.~Zakharov}\affiliation{\landau}

\begin{abstract}
We study the color randomization of two-gluon states produced after 
splitting of a primary fast gluon in the quark-gluon plasma.
We find that 
for the LHC conditions 
the color randomization of the $gg$ pairs is rather slow. 
At jet energies  $E=100$ and $500$ GeV,
for typical jet path length in the plasma in central Pb+Pb collisions, 
the $SU(3)$-multiplet averaged color Casimir of the $gg$ pair
differs considerably from its value $2N_c$ for a fully randomized $gg$ 
state.
Our calculations of the energy dependence for generation of the 
nearly collinear decuplet $gg$ states, that can lead to the baryon
jet fragmentation, show that the contribution of the anomalous decuplet color
states to the baryon production should become small at 
$p_T\gtrsim 10$ GeV. 

\end{abstract}
%

\maketitle
\section{Introduction}
The results of experiments on heavy ion collisions
at RHIC and LHC provide strong evidence for formation 
of a hot quark-gluon plasma (QGP) 
at the proper time $\tau_0\sim 0.5-1$ fm. 
One of the major signals of the QGP formation in $AA$ collisions is
a strong suppression of high-$p_T$ particles (jet
quenching (JQ)) as compared to 
$pp$ collisions. 
The JQ
phenomenon is believed to be a consequence of medium modification of 
the jet fragmentation
functions (FFs) due to radiative \cite{GW,BDMPS,LCPI1,GLV1,AMY,W1} 
and collisional \cite{Bjorken1} parton energy loss in the QGP.
The energy loss is dominated
by the radiative mechanism, and the effect of the collisional
energy loss turns out to be relatively small \cite{Z_coll,Gale_coll}.

The first-principle calculation of the medium modification
of the jet FFs in $AA$ collisions remains an unsolved problem.
The available approaches to the radiative energy loss
\cite{BDMPS,LCPI1,GLV1,AMY} deal with one gluon emission.
In phenomenological applications to the JQ 
multiple gluon emission is usually treated in the approximation of independent
gluon radiation \cite{BDMS_RAA}, similarly to the radiation of soft photons
in QED. This approximation may be reasonable for calculation 
of the nuclear modification factor $R_{AA}$ that is sensitive
mostly to the Sudakov suppression of the FFs at the fractional 
momenta $x$ close to unity. But this may be unsatisfactory
for the soft region $x\ll 1$. 
In principle, the diagram technique of the light-cone path integral
(LCPI) \cite{LCPI1,LCPIpt} approach, originally developed 
for one-gluon emission, 
allows one to go beyond the one gluon level. However, even
at the level of two gluons, and in a crude oscillator approximation
\cite{Z_OA}
(when multiple scattering is described in terms
of the transport coefficient $\hat{q}$ \cite{BDMPS}),
calculations become extremely complicated 
\cite{Arnold_2g1,Arnold_2g2,Arnold_2g3}.
And until now no accurate method has been developed for multiple gluon
emission that could be used for a robust calculation of the in-medium
jet evolution. 
In the last years many efforts have been and are being made on developing 
the Monte-Carlo models of the in-medium parton cascading 
(see e.g. Refs. \cite{W_MC1,PYQUEN,W_MC2,JETSCAPE1}).
These models may be successful in the data analyses but
solid theoretical support for the probabilistic picture,
assumed in the Monte-Carlo schemes, is absent.
  
The problem of the in-medium jet dynamics is complicated by 
the lack of an ordering of scales (say, like the angular ordering
for the vacuum parton cascade) for the induced gluon emission.
It is important that,
in principle, for parton cascade in a finite-size medium 
it is impossible to separate the cases of the ordinary virtuality-ordered 
parton splittings  and that induced by parton rescatterings on the 
medium constituents. For this reason a consistent treatment
should deal with the full medium modified parton cascade.  
At one gluon emission level 
the induced contribution to the gluon spectrum
may be defined as a difference between the spectrum in the medium and
the vacuum one. But this procedure is rather formal, because
this difference includes interference between the vacuum parton 
splitting without interaction with the medium and  
the parton splitting accompanied by parton rescatterings in the medium.
The interference effects are 
important at $L\lsim L_f^{in}$, where $L_f^{in}$
is the typical formation
length for the induced gluon emission in a uniform medium.
Only at large distance from the jet production point,
when the interference terms become small, one may speak
of the purely induced gluon emission.
It is important that for RHIC and LHC conditions,
even for soft gluons with energy $\omega\lsim 3-5$ GeV 
that dominate the induced energy loss,
$L_f^{in}$ may be rather large $\sim 2-5$ fm
\footnote{For an infinite uniform QGP $L^{in}_f\sim 2\omega S_{LPM}/m_g^2$, where
$S_{LPM}$ is the Landau-Pomeranchuk-Migdal suppression factor and
$m_g$ is the gluon quasiparticle mass. 
For RHIC and LHC conditions typically $S_{LPM}\sim 0.3-0.5$ for $m_g\sim 400$ 
MeV \cite{LH}. Then we find that $L_f^{in}\sim 2-5$ fm
at $\omega\sim 3-5$ GeV.}. 
Since this scale is not small as compared to the size
and life-time of the QGP for RHIC and LHC conditions,
we have a situation when inside the QGP the interference 
between the vacuum amplitudes and the ones with rescatterings
is important.
However, if the QGP production time $\tau_0\sim 0.5-1$ fm \cite{Heinz_tau}, 
one can expect that for jets with $E\lsim 100$ GeV
the first most energetic radiated gluon should not be affected strongly, 
because the typical formation length for 
such gluons turns out to be
of the order of (or smaller) than $\tau_0$ \cite{RAA08}. 
But subsequent evolution of the two parton system
produced in the primary parton splitting should be affected by the medium
effects.

It is important that the $t$-channel gluon exchanges between the fast partons 
and the QGP constituents, even for very small momentum kicks, 
can affect the angular-ordered jet evolution.  
One of the mechanisms for this is violation of the color coherence
for the in-medium parton splitting that destroys the angular 
ordering inherent to the vacuum cascade.
In Ref.~\cite{Leonidov1} it was shown that the disruption of the 
angular ordering  leads to a substantial softening 
of the intrajet rapidity spectrum.
For jets with $E\lsim 100$ GeV for which, as was said before, the
formation length for the first parton branching is small,
the medium color decoherence/randomization comes into play for 
gluon emission
from the two-parton states created after decay of the initial hard parton.
One more mechanism for the medium modification of the jet FFs
is connected with change of the total jet color charge
by the $t$-channel gluon exchanges.
The $t$-channel gluons do not change the total color charge for a single 
fast parton. But already after the first in-medium splitting of the 
primary parton, the created two-parton system may belong to
a color multiplet that is impossible for the vacuum cascade
(when for quark and gluon jets the triplet and octet color states
persist for the whole parton cascade).
This is illustrated in Fig.~1
for $g\to gg$ splitting.
The change in the jet color charge may lead to modification
of the jet FFs due to change in the hadronization pattern for fast partons after
escaping from the QGP \cite{W2,AZ_baryon,W3}.  
In Refs.~\cite{W2,W3} this mechanism has been discussed
for $N=1$ and $N=2$ rescatterings
from the point of view of the large-$N_c$ limit
using the cluster and LUND hadronization models. 
It was shown that the medium-modified color flow can contribute to 
the quenching of hadron spectra, and increase the jet FFs
in the soft region. 

Production of the nearly collinear $qg$ systems (for quark jets) 
in the $\{\bar{6}\}$ 
color state and of the $gg$ systems (for gluon jets) 
in the $\{10\}$ and $\{\overline{10}\}$ color states
can lead to an interesting mechanism of the leading
baryon production in jet fragmentation \cite{AZ_baryon,M98}. 
Because after escaping from the QGP
these states may result in creation of color 
tubes with the same anomalous color flux. 
The breaking of these color flux tubes via the Schwinger tunnel $q\bar{q}$ pair
creation produces the color string configurations with the string junction, 
which traces the baryon number in the topological expansion scheme
\cite{RV1,RV2}, that should hadronize into a system  with a leading baryon.
This mechanism of the baryon jet fragmentation for the decuplet $gg$ pairs
is illustrated in Fig.~2 (the interested reader is referred to 
Ref.~\cite{AZ_baryon} for extensive discussion on this mechanism).
An accurate calculation of the contribution of this mechanism to
the baryon production in $AA$ collisions is impossible.
But qualitative analysis performed in Ref.~\cite{AZ_baryon} 
indicates that this mechanism may give a considerable contribution
to the anomalous baryon production at intermediate $p_T$ observed in $AA$ 
collisions at RHIC and LHC \cite{PHENIX_baryon,STAR2,ALICE_baryon}. 
The calculations of Ref.~\cite{AZ_baryon} have been performed
under assumption of a fast randomization of the two-parton states in the QGP.
However, the approximation
of the fully color randomized state becomes invalid for sufficiently high
energies, when the transverse size of the $gg$ pair remains small 
(say, as compared to the Debye radius of the QGP) on the longitudinal
scale about the typical jet path length in the QGP.
To understand better the role of the anomalous color states in 
the baryon jet
fragmentation it would be interesting to perform a quantitative
analysis of the color randomization of two-parton states.

A quantitative analysis of the color randomization of two-parton states
is also of interest in connection with the role of the color decoherence 
in the in-medium  soft gluon  emission by a two-parton antenna that 
depends crucially on the rate of the antenna color randomization.
Over the last years the gluon emission from the two-parton antennas has been
used actively as an interesting theoretical laboratory to 
explore the in-medium multiple gluon emission
(see, e.g. \cite{Iancu_ant1,Iancu_ant2,Salgado_ant1,Salgado_ant2}).
These studies show that, similarly
to the vacuum parton cascade \cite{Basics_QCD}, the coherence effects
are very important in multiple in-medium gluon emission.
Usually the color randomization of the two-parton antenna
is described by a single parameter: the decoherence time 
that characterizes the exponential reduction of the probability
for antenna to stay in the initial color multiplet.
But it would be of interest to see in more detail 
how the color randomization goes. Say, to understand the $L$-dependence
of the distribution of the two-parton state in the irreducible color
multiplets. 
This distribution, e.g.,
is crucial for emission of gluons with the inverse gluon transverse
momentum larger than the transverse size of the two-parton system,
which is only sensitive to the total jet color charge
(both for the vacuum gluon emission \cite{Basics_QCD} from fast partons after
escaping from the QGP
and for the in-medium one \cite{Salgado_2013}). 
The information on the magnitude of the color decoherence
in the QGP  
may also be useful in building of a qualitative picture  
of what goes on with jets in $AA$ collisions in the whole phase space
\cite{Salgado_2013,Kurkela_model,Mueller_model} and for development
of the Monte-Carlo models that account (at least qualitatively) for
the decoherence effects, e.g. like that of Ref.~\cite{Leonidov1}.

In the present paper we perform analysis of the color randomization 
for a $gg$ pair produced from the decay of a primary high energy gluon.
We perform calculation for the QGP corresponding
to conditions of central Pb+Pb collisions at the LHC energy
$\sqrt{s}=2.76$ TeV.
We describe the color randomization with the help of evolution
equation for the color density matrix for the $gg$ system.
The diffraction operator for the four-gluon system, which 
is a crucial ingredient for our calculations, has been previously calculated
in Ref.~\cite{NSZ4g}, devoted to the forward gluon-gluon di-jet production
in $pA$ collisions.
\begin{figure}
\includegraphics[height=2.7cm]{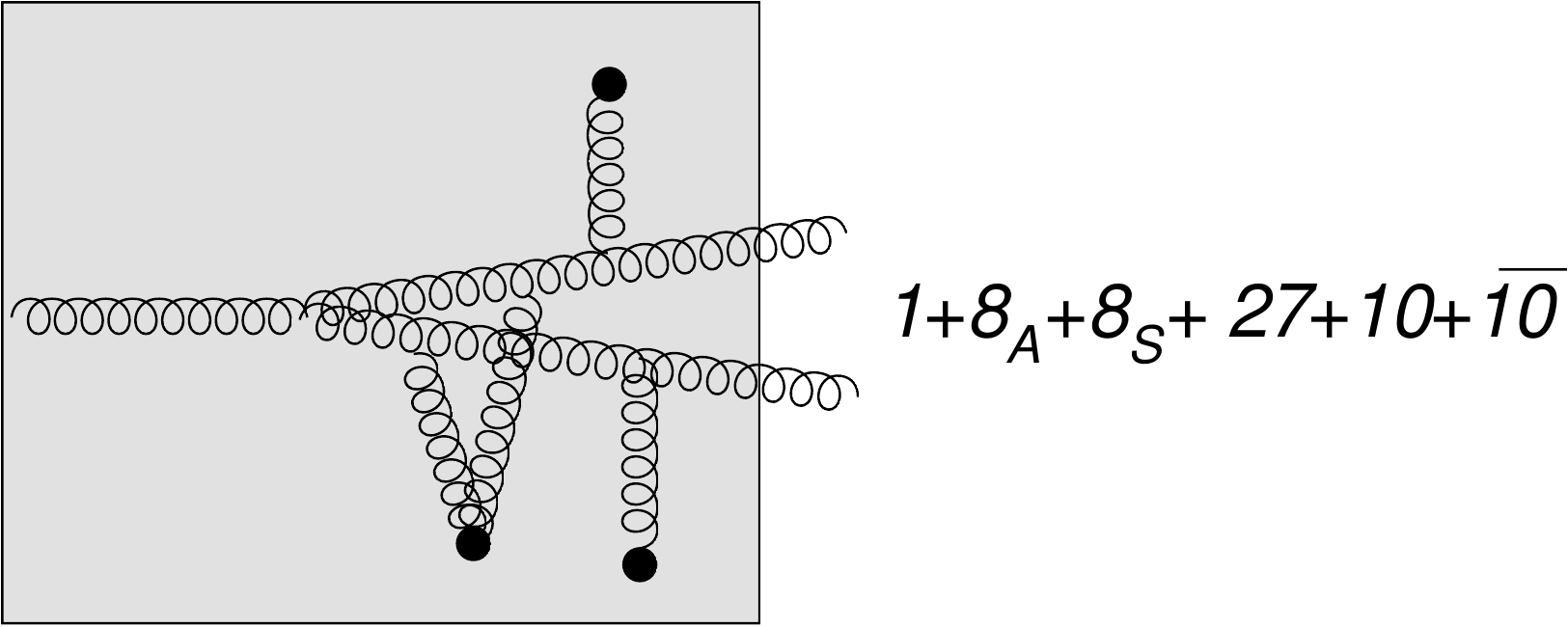}
\caption{\small The $g\to gg$ in-medium splitting and possible color states
of the final two-gluon system.
}
\end{figure}
\begin{figure}
\includegraphics[height=1.7cm]{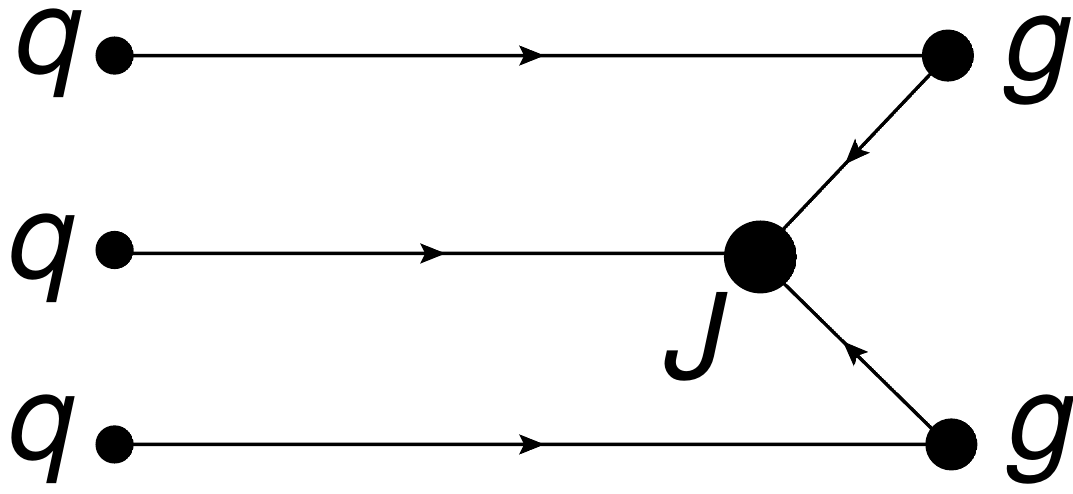}
\hspace{3cm}\includegraphics[height=1.7cm]{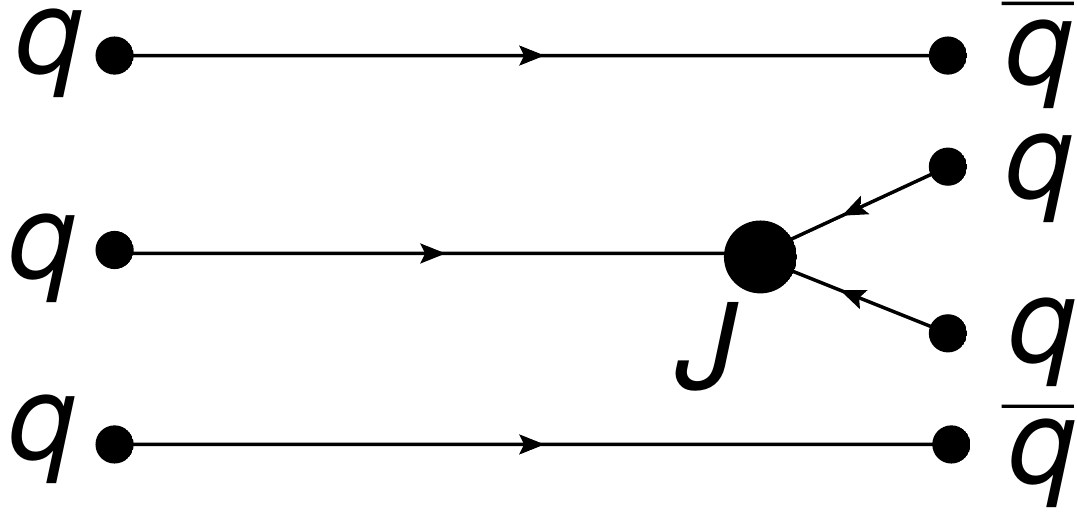}
\caption{\small (left) The color string configuration created by the 
right-moving fast $gg$ pair
in the decuplet color state after escaping from the QGP and color
neutralization of the color flux via the Schwinger production of three 
$q\bar{q}$ pairs, $J$ denotes the string junction \cite{RV1,RV2}. (right)
The same but after splitting of fast gluons into $q\bar{q}$ pairs.
}
\end{figure}

The plan of the paper is as follows. 
In Sec.~2 we describe the formulism for evaluation of the $L$-dependence of
the color density matrix of the $gg$ system in the QGP.
In Sec.~3 we discuss the model of the QGP fireball 
and the parametrization of the dipole cross section
used in our calculations.
In Sec.~4  we present the numerical results. 
We give conclusions in Sec.~5.
Some formulas relevant to our calculations are given  in Appendix.

\begin{figure}
\includegraphics[height=3.1cm]{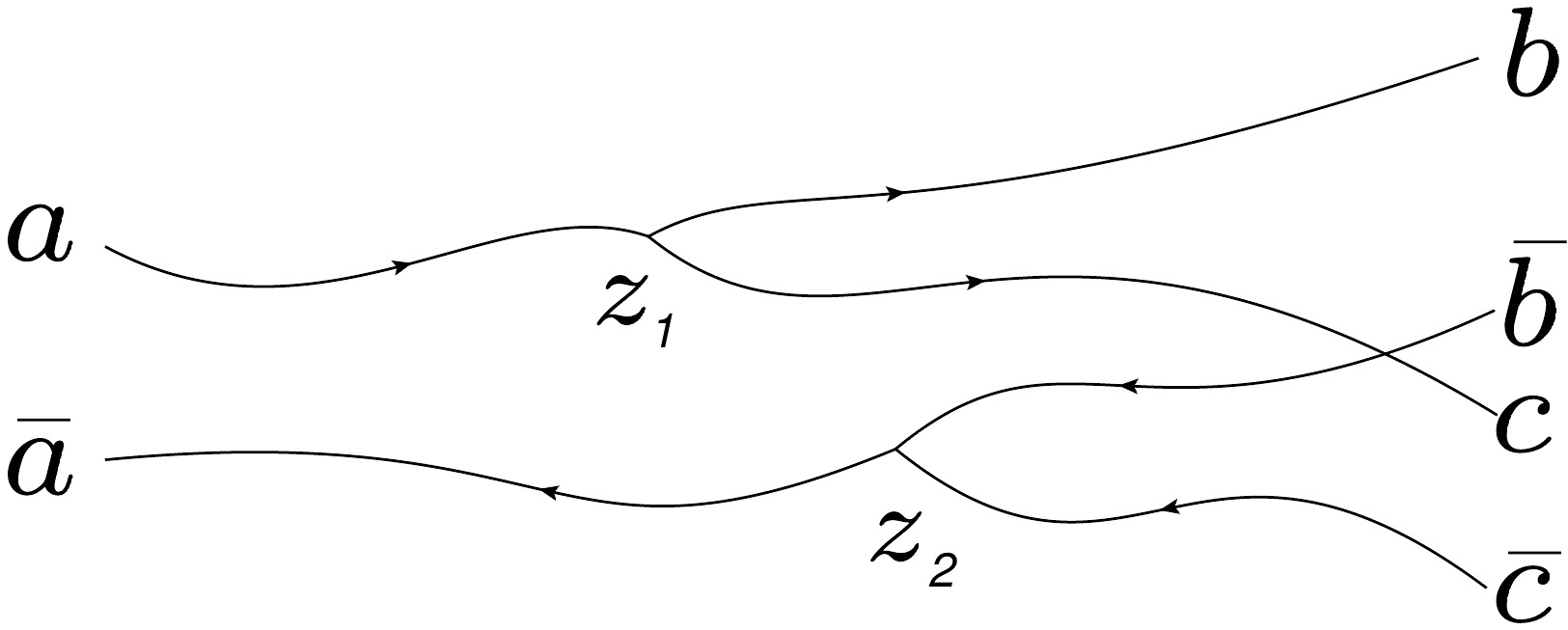}
\caption{\small Schematic diagram picture of the light-path
integral representation for the squared amplitude 
$|\langle bc|T|a\rangle|^2$ describing
the probability of $a\to bc$ transition
in the LCPI \cite{LCPIpt} approach.
The lines with right and left arrows correspond to the amplitude and 
complex conjugate amplitude, respectively. For the in-medium
transition the lines can interact with the medium constituents
via the $t$-channel gluon exchanges.
}
\end{figure}
\begin{figure}
\includegraphics[height=2.7cm]{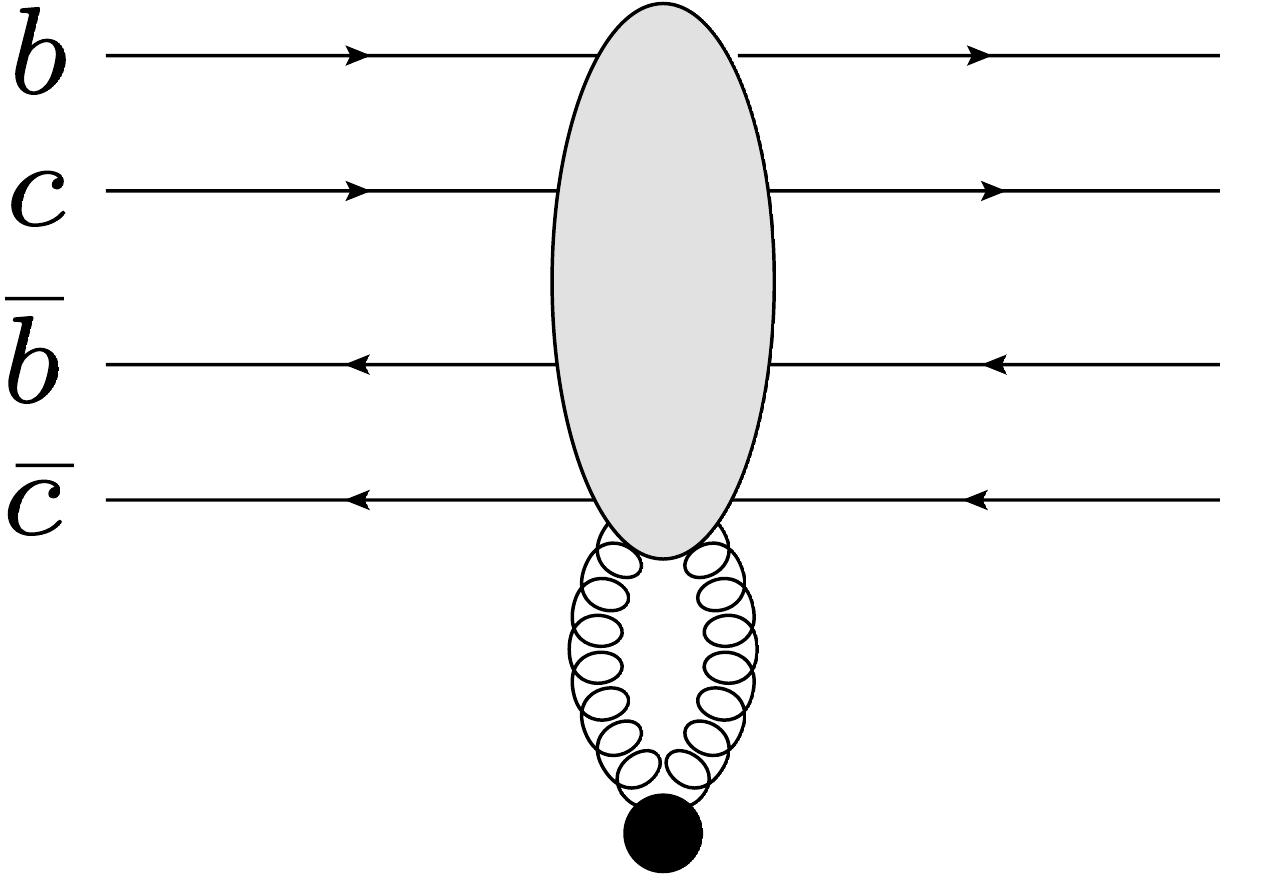}
\caption{\small The diffractive operator for scattering of 
the four-body $bc\bar{b}\bar{c}$ system on a medium constituent in 
the two-gluon 
approximation. The blob includes all possible attachments of the $t$-channel 
gluons
to the parton and antiparton lines.
}
\end{figure}
\begin{figure}
\includegraphics[height=2.7cm]{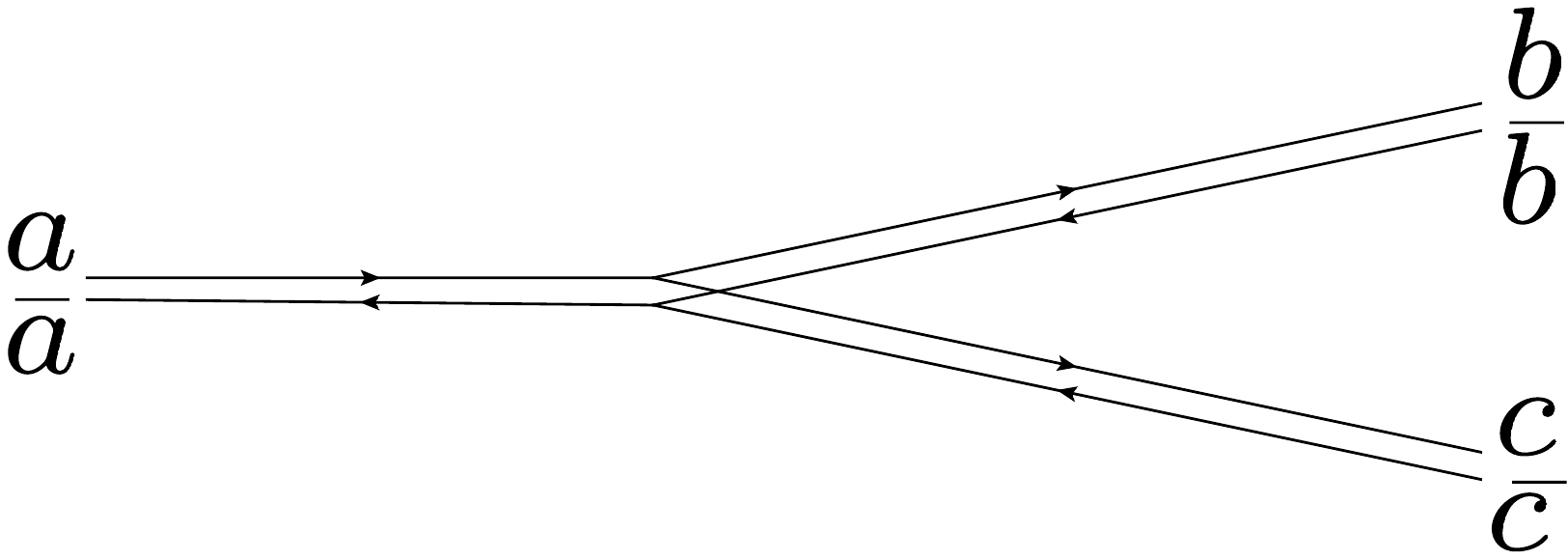}
\caption{\small 
The parton trajectories 
for the squared amplitude 
$|\langle bc|T|a\rangle|^2$ for $a\to bc$ splitting
in the approximation of rigid
geometry used in the present analysis.
}
\end{figure}

\section{Evolution of the color density matrix of $gg$ pair}
In this section we formulate our model for in-medium evolution of the 
color state of a two-gluon pair produced via splitting 
$g\to gg$ of an initial hard gluon with energy $E$. We assume that
the parent fast gluon is produced at $z=0$ (we choose $z$-axis along
its momentum, so $z$ equals the jet path length $L$ in the QGP). 
We will describe the QGP in the approximation of static color Debye-screened
scattering centers \cite{GW}.

In general, in the LCPI formalism \cite{LCPIpt} 
the probability of $a\to bc$
splitting in the small angle approximation may be represented 
as a path integral over the transverse 
parton coordinates on the light-cone $t-z=$const shown diagrammatically
in Fig.~3, where the right and left arrows denote the trajectories for the
amplitude and the complex conjugate amplitude, respectively.
The color generators for a parton $p$ and for its 
antipartner $\bar{p}$ satisfy the relation
$T^{\alpha}_{\bar{p}}=-(T^{\alpha}_p)^{*}$ (for $p=g$ we have $\bar{g}=g$).
With the help of this relation one can show that, 
in the approximation of two-gluon exchanges between fast partons and 
medium constituents, the parton lines corresponding to the complex 
conjugate amplitude interact with the medium similarly to 
the antiparton lines. It means that in the
path integral the interaction part of the Lagrangian is analogous to that for
a fictitious system of partons (upper lines in Fig.~3) and antipartons
(lower lines in Fig.~3). This system is fictitious
because in the Lagrangian the kinetic term  for antipartons is negative
due to complex conjugation. 
It is important that the  fictitious parton-antiparton   system 
at any $z$ is in a color singlet state. 
Indeed, since we perform averaging over the color states 
of the initial hard parton $a$, at the initial instant $z=0$ 
we have the $a\bar{a}$ pair in the color singlet state
(this means that the two gluon lines at the initial instant   
in Fig.~3 become closed in the sense of the color flow.)
And the subsequent $t$-channel two-gluon exchanges do not change the total 
color charge of our fictitious system. It occurs because only color singlet
two-gluon states survive after summing over the final states of the medium
with the help of the closure relation. After this operation is done
at the level of the integrand, 
the effect of the $t$-channel gluon exchanges
\footnote{In the literature, the interaction of parton trajectories
with QCD matter is often described in terms of the Wilson line 
factors. This may 
create an impression that the picture with the fictitious color 
singlet parton-antiparton system interacting with the medium is valid 
even for nonperturbative fluctuation of the color fields of the medium.
But this is not the case, because for nonperturbative situation
the vector potentials in the Wilson lines for the amplitude 
and in the ones for the complex conjugate amplitude may be different.
Even in the perturbation theory the validity of this picture
is limited only to the two-gluon $t$-channel exchanges.}
 translates to appearance in the Lagrangian for the fictitious parton-antiparton
system interaction between the trajectories described 
by an imaginary potential
\beq
V(z,\{\ro\},\{\bar{\ro}\})
=-i\frac{n(z)\hat{\sigma}(z,\{\ro\},\{\bar{\ro}\})}{2}\,,
\label{eq:10}
\eeq
where $\{\ro\}$ and $\{\bar{\ro}\}$ are the sets of the transverse 
coordinates of the partons and antipartons,   
$n(z)$ is the number density of the medium constituents,
and $\hat{\sigma}(z,\{\ro\},\{\bar{\ro}\})$ is the diffraction
operator for scattering of the fictitious system on the medium
constituent via the two-gluon exchanges (as shown in Fig.~4
for the four-parton system $bc\bar{b}\bar{c}$). In (\ref{eq:10}) (and below) 
for notational simplicity we omit the sum over species of the medium
constituents (quarks and gluons). 
In the diagram of Fig.~3 for the two-body ($z<z_1$) and three-body 
($z_1<z<z_2$) parton-antiparton systems only one color singlet state is possible
\footnote{For three gluons there are two
color singlet states: asymmetric $\propto f_{\alpha\beta\gamma}$ and symmetric
$\propto d_{\alpha\beta\gamma}$. However, in the case of the $g\to gg$ 
splitting the three-body system in the diagram of Fig.~3
may be only in asymmetric color state, because after 
$g\to gg$ transition two gluons are in asymmetric color octet state,
and the $t$-channel gluon exchanges cannot change the symmetry 
of the three-gluon color wave function.}.
In both the cases, the color singlet states are the eigenstates of 
the diffraction operator, for this reason for the two- and three-body
parts the diffraction operator may simply be replaced 
by the total cross sections for the two- and three-body
color singlet systems. For the four-body part at $z>z_2$ the situation 
is more complicated, because there are several color singlet four-parton 
states, and the diffraction operator has off-diagonal 
elements between them. For this reason for the four-body part in the
diagram of Fig.~3
the path integration over the transverse coordinates 
cannot be separated from the color algebra.
 
In the present work we perform calculations using 
the rigid geometry with 
the straight trajectories, same for the amplitude and  the complex conjugate
one as shown in Fig.~5 for $a\to bc$ transition. 
This approximations seems to be reasonable
for relatively hard parton splittings, when fluctuations of
the trajectories of energetic partons are small and the separation
between the jet production point for the amplitude and the complex conjugate
amplitude ($\sim 1/Q$) is much smaller than the typical scale for 
color fields in the QGP, say than the Debye radius.
The approximation of straight trajectories for fast partons has been
widely used for analysis of soft gluon emission in hard QCD process.
E.g. this picture
has been used in calculations of the anomalous dimension matrix
for large angle soft gluon emission in hard $gg\to gg$ scattering
\cite{hardQCD1,hardQCD2,DM1,hardQCD3}.  One can say that the approximation
of straight lines, in terminology of the recent analysis
\cite{Mueller_model}, should be reasonable for the vacuum-like
emission. However, we will also apply it  
to a relatively soft $g\to gg$ splitting (where its applicability is 
questionable)  for a qualitative analysis of the color decoherence 
in the induced gluon emission.

In our case of $g\to gg$ transition the fictitious color singlet four-body
$bc\bar{b}\bar{c}$ system is simply a color singlet four-gluon system
because gluon is a self-conjugate particle
and $gg\bar{g}\bar{g}=gggg$.
We will label the final two gluons in the amplitude $g_1$ and $g_2$,
and the final two gluons in the complex conjugate amplitude  $g_3$ and $g_4$.   
We describe the color state of the two gluon system 
by the density matrix 
$\langle a b |\hat{\rho}|c d\rangle$\,,
where $a$, $b$, $c$ and $d$ refer to the color indexes
of the gluons $g_1$, $g_2$, $g_3$ and $g_4$, respectively
\footnote{
Note that we consider the situation when the color indexes for final gluons in
the amplitude and the complex conjugate amplitude may differ. 
This differs from calculation
of the gluon spectrum,  when one performs summing over $a=c$ and $b=d$ 
\cite{LCPIpt,Blaizot1,Salgado_coh}, and
the final gluon lines 
with right and left arrows in Figs.~3,~5 become closed (in the sense of the
color flows).}
The imaginary potential (\ref{eq:10}) in the path integral formulation
corresponds to the evolution equation for the two-gluon color density matrix 
given by
\beq
\frac{d\hat{\rho}(z)}{dz}=-\frac{n(z)}{2}\hat{\sigma}\hat{\rho}\,,
\label{eq:20}
\eeq
where 
$\hat{\sigma}$ is the diffraction operator for scattering of the 
four-gluon system due to the double gluon exchange
as shown in Fig.~4. 
The color density matrix of 
the two-gluon system in
(\ref{eq:20}) may be viewed as the color wave function of the fictitious 
color singlet four-gluon system
\beq
\langle ab cd|\Psi\rangle=\langle ab|\hat{\rho}|cd\rangle\,.
\label{eq:30}
\eeq

The color wave functions of the gluon pairs $g_1g_2$ and $g_3g_4$ 
may belong to one of the irreducible multiplets
in the Clebsch-Gordan decomposition of the direct product of two octets 
\beq
8\otimes 8=1+8_A+8_S+27+10+\overline{10}\,,
\label{eq:40}
\eeq   
where $8_A$ and $8_S$ denote the antisymmetric and symmetric octet
states that may be built from the $SU(3)$ tensors $f_{\alpha\beta\gamma}$
and $d_{\alpha\beta\gamma}$, respectively.
From the irreducible multiplets in the Clebsch-Gordan decomposition
(\ref{eq:40}) one can build eight color singlet states for 
the four-gluon system. There are six states of the types
$|RR\rangle$/$|R\bar{R}\rangle$ 
\beq
|11\rangle,\,\,\,|8_A8_A\rangle,\,\,\, |8_S8_S\rangle,\,\,\,
|27\,27\rangle,\,\,\,|10\,\overline{10}\rangle,\,\,\,
|\overline{10}\,10\rangle,\,\,\,
\label{eq:50}   
\eeq
and two mixed states built from different octet multiplets
\beq
|8_A8_S\rangle,\,\,\, |8_S8_A\rangle\,.
\label{eq:60}   
\eeq
The in-medium color randomization of the two-gluon system produced
in the process $g\to gg$ can be described in terms of the six states given 
in (\ref{eq:50}) (in our formulas we will denote these states as 
$|R\bar{R}\rangle$ even for self-conjugate multiplets when 
$|R\bar{R}\rangle=|RR\rangle$).
The point is that the initial two-gluon state for 
the transition $g\to gg$ is the antisymmetric octet $8_A$.
In terms of the description of the density matrix via 
the four-gluon wave function 
it corresponds to the color singlet $|8_A8_A\rangle$. 
The subsequent in-medium evolution of this state (and of any other state
of the types $|R\bar{R}\rangle/|RR\rangle$) cannot
generate the mixed states (\ref{eq:60}) because the matrix elements
of the diffraction operator $\hat{\sigma}$ between 
the states from (\ref{eq:50}) and (\ref{eq:60}) vanish. This means 
that the mixed
states (\ref{eq:60}) turn out to be fully decoupled from the color randomization of the
two gluon system.
Thus, the in-medium four-gluon wave function can be decomposed
as a sum over the singlet color states given in (\ref{eq:50}) 
\beq
\langle ab cd|\Psi\rangle=\sum_{R}c_R
\langle ab cd|R\bar{R}\rangle\,.
\label{eq:70}   
\eeq
The corresponding decomposition of the density matrix
can be written as 
\beq
\langle ab|\hat{\rho}|cd\rangle
=\sum_{R}P_R \langle ab|\hat{\rho}_R|cd\rangle\,,
\label{eq:80}   
\eeq
where $P_R$ is the probability that the two-gluon system 
belongs to the multiplet $R$, and $\hat{\rho}_R$ is the 
density matrix for the multiplet $R$.
The full density matrix $\hat{\rho}$ and its components $\hat{\rho}_R$ satisfy the normalization conditions
\beq
\sum_{a,b}\langle ab|\hat{\rho}|ab\rangle=1\,,\,\,\,\,\,   \sum_{a,b}\langle ab|\hat{\rho}_R|ab\rangle=1\,.
\label{eq:90}   
\eeq

The $z$-dependence of the vector
\beq
\vec{P}=(P_{1},P_{8_A},P_{8_S},P_{27},P_{10},P_{\overline{10}})
\label{eq:100}   
\eeq
characterizes the process of the in-medium color randomization of the 
$gg$ pair. 
The in-medium evolution should satisfy the conservation 
of the total probability to find the $gg$ pair in any color
state 
\beq 
\sum_{R}P_R=1\,.
\label{eq:110}   
\eeq
In the limit of very large medium thickness the $gg$ pair should tend to the
fully color randomized state,
when $P_R$ is defined by the multiplet dimensions
\beq 
\left.P_R\right|_{\rm randomized}=\frac{\mbox{dim}[R]}{\sum_{R'}\mbox{dim}[R']}
=\frac{\mbox{dim}[R]}{(N_c^2-1)^2}\,.
\label{eq:120}   
\eeq

The normalized to unity density matrix $\hat{\rho}_R$ for a given 
multiplet $R$ can be written as
\beq
\langle ab|\hat{\rho}_R|ab\rangle=
\frac{1}{\mbox{dim}[R]}P[R]^{ab}_{cd}\,,
\label{eq:130}   
\eeq
where
\beq
P[R]^{ab}_{cd}=\sum_{\nu}\langle ab|R\nu\rangle\langle R\nu|cd\rangle
\label{eq:140}   
\eeq
is the projector onto the states of the irreducible multiplet $R$
(here $\nu$ labels the states in the multiplet $R$).
The fact that $\hat{\rho}_R$ given in (\ref{eq:130}) is normalized to unity 
is a consequence of the relation 
\beq
\sum_{a,b}P[R]^{ab}_{ab}=\mbox{dim}[R]\,.
\label{eq:150}   
\eeq
The derivation of the formulas for the projectors can be found in
Refs. \cite{DM1,NSZ4g}. 
For the reader's convenience we present them in Appendix.

The four-gluon color wave function $\langle ab cd|R\bar{R}\rangle$
in terms of the projector $P[R]^{ab}_{cd}$ reads
\beq
\langle ab cd|R\bar{R}\rangle=\frac{1}
{\sqrt{\mbox{dim}[R]}}P[R]^{ab}_{cd}\,.
\label{eq:160}   
\eeq
From the fact that $P[R]P[R]=P[R]$ and the relation (\ref{eq:150}) one can see
that the wave function (\ref{eq:160}) 
is normalized to unity, i.e.
\beq
\sum_{abcd}\langle R\bar{R}|abcd\rangle\langle abcd| R\bar{R}\rangle=1\,.
\label{eq:170}   
\eeq
We explicitely show here the sum over the gluon color states to
demonstrate that the normalizations conditions for the components 
of the density
matrix for a given multiplet $\langle ab|\hat{\rho}_R|cd\rangle$  and the color singlet four-gluon wave
function $\langle abcd| R\bar{R}\rangle$ built from $R$ and $\bar{R}$ are defined in different ways.

From (\ref{eq:20}) and (\ref{eq:30}) one can easily obtain the 
evolution equation
in terms of the coefficients $c_R$ in the four-gluon wave function decomposition
into the $R\bar{R}$ states (\ref{eq:70}) 
\beq
\frac{dc_R}{dz}=-\frac{n(z)}{2}
\langle R\bar{R}|\hat{\sigma}|R'\bar{R}'\rangle c_{R'}\,.
\label{eq:180}   
\eeq
The formulas for the diffraction operator in the basis of the color singlet
states $|R\bar{R}\rangle$ for arbitrary gluon positions have been derived in
\cite{NSZ4g}. 
For the reader's convenience in Appendix we present the formula
for the diffraction matrix $\langle R\bar{R}|\hat{\sigma}|R'\bar{R}'\rangle$
for the gluon configurations for the rigid geometry (as shown in Fig.~5)
with the transverse coordinates $\bb_1=\bb_3$ and $\bb_2=\bb_4$,
that is used  in the present analysis. 
Note that the off-diagonal elements 
of the diffraction operator are nonzero  
only between the multiplets with different permutation
symmetry.
From (\ref{eq:130}), (\ref{eq:160}) one can see that the relation between   
the coefficients $c_R$ in the decomposition of the wave function 
(\ref{eq:70}) and the 
coefficients $P_R$
in the decomposition
of the density matrix (\ref{eq:80}) reads 
\beq
c_R=P_R/\sqrt{\mbox{dim}[R]}\,.
\label{eq:190}   
\eeq
Then, the evolution equation (\ref{eq:180})  
in terms of the coefficients $P_R$ can be written as
\beq
\frac{dP_R}{dz}=-\frac{n(z)}{2}
\langle R\bar{R}|\hat{\sigma}|R'\bar{R}'\rangle P_{R'}
\sqrt{\frac{\mbox{dim}[R]}{\mbox{dim}[R']}}\,.
\label{eq:200}   
\eeq

It worth noting that, in the description of the in-medium
evolution in terms of the four-gluon wave function,
$\langle\Psi|\Psi\rangle =\sum_R |c_R|^2$ does not 
corresponds to the total probability to find the two-gluon system 
in any color state.  And the evolution equation (\ref{eq:180}) 
does not conserve $\sum_R |c_R|^2$.  
The total probability to find the two-gluon system in any color state
is given by the sum $\sum_R P_R$. The fact the evolution
equation (\ref{eq:200}) preserves the probability conservation 
 for $P_R$ (\ref{eq:110}) is a non-trivial 
consequence of the color transparency
for the point-like color singlet parton states.
Indeed, from (\ref{eq:200}) one can see that the conservation 
of the sum $\sum_R P_R$   requires fulfilling
the relation
\beq
\sum_R\langle R\bar{R}|\hat{\sigma}|R'\bar{R}'\rangle
\sqrt{\mbox{dim}[R]}=0
\label{eq:210}   
\eeq
for any $R'$. One can write the left-hand side of (\ref{eq:210}) as
\beq
\sum_R\langle R\bar{R}|abcd\rangle\langle abcd|\hat{\sigma}|R'\bar{R}'\rangle
\sqrt{\mbox{dim}[R]}=
\sum_R P[R]^{ab}_{cd}\langle abcd|\hat{\sigma}|R'\bar{R}'\rangle\,.
\label{eq:220}   
\eeq
But from the closure relation for the projectors
\beq
\sum_R P[R]^{ab}_{cd}=\openone^{ab}_{cd}
=\delta_{ac}\delta_{bd}\,,
\label{eq:230}   
\eeq
 one obtains for the left-hand
side of (\ref{eq:220})
\beq
\sum_{ab}\langle abab|\hat{\sigma}|R'\bar{R}'\rangle
\propto \langle (g_1g_3)_{\{1\}}(g_2g_4)_{\{1\}}
|\hat{\sigma}|R'\bar{R}'\rangle\,,
\label{eq:240}   
\eeq
where $(g_1g_3)_{\{1\}}$ and $(g_2g_4)_{\{1\}}$ denote the color singlet
states of the gluon pairs $g_1g_3$ and $g_2g_4$. In our case 
these pairs have a zero size. For such configurations the total contribution
of the $t$-channel two-gluon exchanges should vanish. This proves the
conservation of the sum $\sum_R P_R$.
Note that
the above consideration also works to prove that for the vector (\ref{eq:100})
with $P_R$ for the regime of complete color randomization defined
by (\ref{eq:120}) the right-hand side of (\ref{eq:200}) vanishes
(it means that 
(\ref{eq:120}), in terms of the coefficients $c_R$, 
corresponds to the eigenvector of the diffraction operator with zero
eigenvalue).
Indeed, for $P_{R'}$ defined by (\ref{eq:120}) the right-hand side 
of (\ref{eq:200}) is proportional to
\beq
\sum_{R'}\langle R\bar{R}|\hat{\sigma}|R'\bar{R}'\rangle
\sqrt{\mbox{dim}[R']}
\label{eq:250}   
\eeq
that, similarly to (\ref{eq:210}), vanishes for any multiplet $R$.

The above formulas correspond to the color singlet four-gluon
 states written in terms of the color states of the gluon pairs $g_1g_2$ and
$g_3g_4$. But one can describe the four-gluon
system in terms of the color singlet states constructed from 
the pairs $g_1g_3$ and $g_2g_4$. As in Ref. \cite{NSZ4g} we 
call these two bases the $s$- and $t$-channel bases.
The $t$-channel basis states can be obtained from the $s$-channel ones
by a unitary transformation $U_{ts}$ (and the inverse matrix $U_{st}$ 
transforms the $t$-channel states into the $s$-channel states). 
Since the complete set of the color 
singlet four-gluon
states includes the mixed states (\ref{eq:60}), the dimension of the 
crossing matrix
$U_{ts}$ is $8\times8$. The matrix $U_{ts}$ was calculated in 
\cite{NSZ4g}. For the reader's convenience we present $U_{ts}$ in 
Appendix (we correct some misprints in the formula C17 of Ref.~\cite{NSZ4g}).
It is convenient to write the crossing matrix using
for the mixed states (\ref{eq:60}) the linear combinations
\beq
|(8_A8_S)_{\pm}\rangle=\frac{i}{\sqrt{2}}\left(
|8_A8_S\rangle\pm|8_S8_A\rangle\right)\,.
\label{eq:260}   
\eeq
In this basis the $t$-channel state $|(8_A8_S)^{-}\rangle$
has a non-zero projection only on the same $s$-channel 
state $|(8_A8_S)^{-}\rangle$, and the state $|(8_A8_S)^{+}\rangle$ in the
$t$-channel basis
has non-zero projections only on the states 
$|10\overline{10}\rangle$ and $|\overline{10}10\rangle$ in 
the $s$-channel basis.
By appropriate choice of the phase factors for
the  $|8_A8_S\rangle$ and $|8_S8_A\rangle$ states,
the unitary crossing matrix can be made real and symmetric, i.e.,
we have $U_{ts}=U_{st}$ and $U_{ts}^2=\openone$ (see Appendix for details).
Note that this is possible only for the complete set of the 
color singlet states, i.e., for the $8\times 8$ crossing matrix.

The solution of the evolution equation (\ref{eq:180}) can be expressed via
the eigenfunctions of the diffraction matrix. The eigenvectors can
be easily written in terms of the $t$-channel states analogous to (\ref{eq:50}) 
and the linear combinations (\ref{eq:260}) of the $8_A8_S$ and $8_S8_A$ states.  
The point is that $g_1g_3$ and $g_2g_4$ pairs are the point-like
objects. For this reason the $t$-channel gluons cannot resolve their internal
color structure, and do not change the color
multiplets for $g_1g_3$ and $g_2g_4$ pairs. As a result, the diffraction
operator in the $t$-channel basis $|\Psi_i^t\rangle$   has a simple
diagonal form with
\beq
\langle \Psi_i^t|\hat{\sigma}|\Psi_j^t\rangle
=\delta_{ij}\sigma_{R_i}(\rho_{12})\,,
\label{eq:270}   
\eeq
where   $R_i$ denotes the color multiplet the state $\Psi_i^t$
is built from, and 
$\sigma_{R_i}$ is the dipole cross section for the color singlet
state $R_i\bar{R}_i$ (from the point of view of the dipole
cross section there is no difference between $8_A$ and $8_S$ octets),
$\rho_{12}=|\bb_1-\bb_2|$ is the transverse size of the  $g_1g_2$ pair
(which equals to that for the  $g_3g_4$ pair). 
In the approximation of the static Debye-screened
scattering centers  \cite{GW} the dipole cross section for the color 
singlet $R\bar{R}$ 
state reads
\beq
\sigma_{R}(\rho)=C_{T}C_{R}\int d\qbt \alpha_{s}^{2}(\qbt^2)
\frac{[1-\exp(i\qbt\ro)]}{(\qbt^{2}+m_{D}^{2})^{2}}\,,
\label{eq:280}
\eeq
where $m_D$ is the Debye mass, $C_T$ and $C_R$ are
the color Casimir operators for the QGP constituent and the multiplet $R$.
The $SU(3)$ Casimir operators that we need read: 
$C_1=0$, $C_8=N_c$, $C_{10}=2N_c$,
$C_{27}=2(N_c+1)$. 
The six eigenstates of the diffraction operator 
in the $s$-channel basis can be obtained by acting with the crossing
matrix $U_{st}$ on the $t$-channel states
\beq
|11\rangle_t,\,\,|8_A8_A\rangle_t,\,\, |8_S8_S\rangle_t,\,\, 
|(8_A 8_S)_{+}\rangle_t,\,\, |27\,27\rangle_t,\,\,
(|10\overline{10}\rangle_t+|\overline{10}10\rangle_t)/\sqrt{2}\,.
\label{eq:290}   
\eeq

For each eigenstate the medium effect 
is reduced to trivial multiplication by the Glauber attenuation factor 
\beq
S_R(z,z_s)=\exp\left[-\frac{1}{2}\int_{z_s}^{z}dz
n(z)\sigma_{R}(\rho_{12}(z))\right]\,,
\label{eq:300}   
\eeq 
where $z_s$ is the longitudinal coordinate of the $g\to gg$ splitting,  $R$ is the multiplet entering the $t$-channel state.
The eigenstate corresponding to the state 
$|(8_A 8_S)_{+}\rangle_t$ in terms of the $s$-channel basis is
$(|10\overline{10}\rangle-|\overline{10}10\rangle)/\sqrt{2}$,
i.e., it describes the difference between the probabilities
for the decuplet $P_{10}$ and the antidecuplet states $P_{\overline{10}}$.
Since for the $g\to gg$ splitting at initial instant 
$P_{10}=P_{\overline{10}}$, one can simply ignore this state. 
Then, the $z$-dependence of the coefficients  $c_R$
in the $s$-channel basis can written as
\beq
c_R(z)=\sum_{R',R_0}\langle R\bar{R}|U_{st}|R'\bar{R}'\rangle
S_{R'}(z,z_s)\langle R'\bar{R}'|U_{ts}|R_0\bar{R}_0\rangle
c_{R_0}(z_s)\,,
\label{eq:310}   
\eeq 
where sum over the intermediate $t$-channel states includes only the first
six states of the type $|R\bar{R}\rangle$.
Thus, for $g\to gg$ splitting we need only $6\times 6$ block of the full
$8\times 8$ crossing matrix.
From (\ref{eq:310}) we obtain for the vector $\vec{P}$ 
\beq
P_R(z)=\sum_{R',R_0}\langle R\bar{R}|U_{st}|R'\bar{R}'\rangle
S_{R'}(z,z_s)\langle R'\bar{R}'|U_{ts}|R_0\bar{R}_0\rangle
P_{R_0}(z_s)\sqrt{\frac{\mbox{dim}[R]}{\mbox{dim}[R_0]}}\,.
\label{eq:320}   
\eeq

In the limit of very large thickness in (\ref{eq:310}), (\ref{eq:320}) 
in the sum over the 
intermediate $t$-channel states only the $|11\rangle$ state 
with $\sigma_{1}=0$  and $S_1=1$ survives, and one can write
(\ref{eq:320}) at $z\to \infty$ as 
\beq
\left.P_R(z)\right|_{z\to \infty}\approx
\sum_{R_0}\langle R\bar{R}|U_{st}|11\rangle
\langle 11|U_{ts}|R_0\bar{R}_0\rangle
P_{R_0}(z_s)\sqrt{\frac{\mbox{dim}[R]}{\mbox{dim}[R_0]}}\,.
\label{eq:330}   
\eeq 
The $t$-channel color singlet wave function
is given by $\langle abcd|11\rangle=\delta_{ac}\delta_{bd}/(N_c-1)^2$.
Using this formula  with the help of (\ref{eq:150}) and (\ref{eq:160}) 
one obtains 
$\langle R\bar{R}|U_{st}|11\rangle=\sqrt{\mbox{dim}[R]}/(N_c^2-1)$,
and a similar formula for $\langle 11|U_{ts}|R_0\bar{R}_0\rangle$. 
Then, with these matrix elements the right-hand side of (\ref{eq:330}) 
is reduced to the color randomized distribution (\ref{eq:120}).

For the initial two-gluon state produced via the gluon
splitting $g\to gg$ in the right-hand side of (\ref{eq:310}) and (\ref{eq:320}) 
there is no summing over $R_0$ because in this case
we have only one non-zero component for $R_0=8_A$  with $P_{8_A}=1$.

\section{Model of the QGP fireball and the dipole cross section}
We perform numerical calculations for the model of the QGP fireball 
with Bjorken's  1+1D expansion \cite{Bjorken}, that, for the ideal
gas model, gives 
$T_{0}^{3}\tau_{0}=T^{3}\tau$, where $\tau_0$
is the thermalization time of the matter.  
As in our previous analyses of the JQ phenomenon 
\cite{RAA08,RAA13,RPP14}, we take $\tau_{0}=0.5$ fm, and for simplicity 
we neglect variation of
$T_{0}$ with the transverse coordinates.
To account for the fact that
the QGP formation is clearly not an instantaneous process,
we take the medium density 
$\propto \tau$ at $\tau<\tau_{0}$.
We fix the initial QGP temperature from the initial entropy density determined 
via the charged particle multiplicity pseudorapidity density,
$dN_{ch}^{AA}/d\eta$, at mid-rapidity  ($\eta=0$) with the help of the 
Bjorken relation \cite{Bjorken}
\beq
s_{0}=\frac{C}{\tau_{0} S_{f}}\frac{dN_{ch}^{AA}}{d\eta}\,.
\label{eq:340}
\eeq
Here $C=dS/dy{\Big/}dN_{ch}^{AA}/d\eta\approx 7.67$ \cite{BM-entropy} 
is the entropy/multiplicity ratio,
and $S_{f}$ is the transverse area of the QGP fireball. 
For the central Pb+Pb collisions at $\sqrt{s}=2.76$ TeV
this procedure for the ideal gas model gives $T_0\approx 420$ MeV
(we take $N_{f}=2.5$ to account for the mass suppression for the strange
quarks in the QGP).
For the above value of $T_0$ in $1+1D$ Bjorken's expansion
the QGP reaches $T\sim T_c$ (here $T_c\approx 160$ is the crossover
temperature) at $\tau_{QGP}\sim 10$ fm. This means that the matter
remains in the plasma phase until a strong cooling of the matter
at $\tau\gsim (1-2)R_A$ 
(here $R_A\sim 6$ fm is the nucleus radius),
when the transverse expansion becomes very strong \cite{Bjorken}.   

\begin{figure}
\includegraphics[height=5cm]{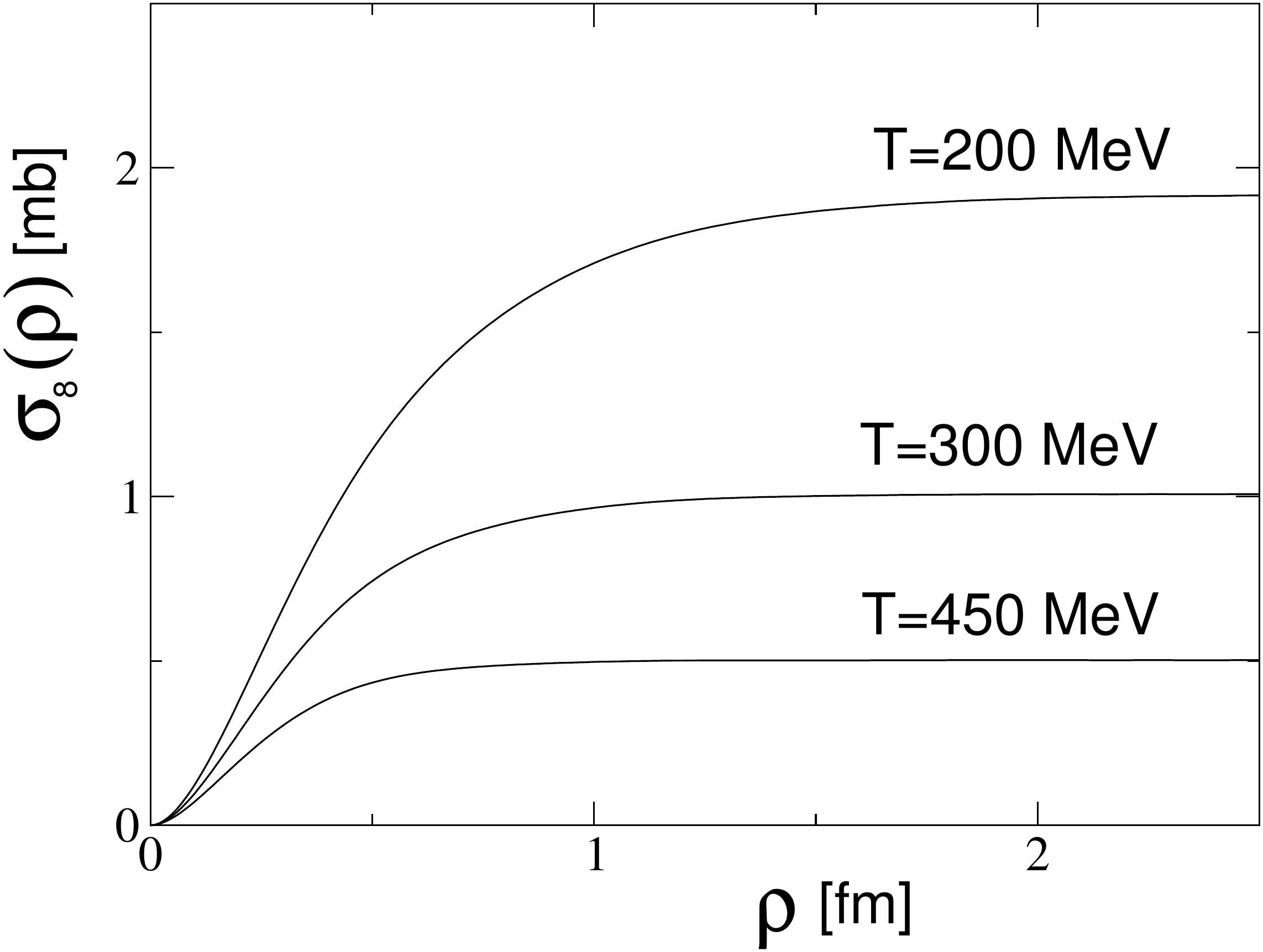}
\caption{\small The octet-octet dipole cross section on a quark
in the QGP versus the dipole size $\rho$ for different values of the 
QGP temperature $T$.
}
\end{figure}
In our calculations the properties of the QGP enter only through the
product of the number density of the QGP and the dipole cross section 
in the formula for the Glauber attenuation factors (\ref{eq:300}). 
Since the dipole cross section is proportional to the Casimir operator 
of the scattering center, in this product one can
avoid summing over the species of the QGP constituents by
using the dipole cross section for scattering on a quark and  
using, at the same time, for the number density of the color centers
the sum $n=n_q+n_gC_A/C_F$ (here $n_q$ is the number density of quarks and 
antiquarks, and $n_g$ is the number density of gluons, $C_A$ and $C_F$
are the gluon and quark Casimir operators).
In calculating the dipole cross section we use 
the Debye mass $m_D$ in the QGP obtained in the lattice 
analysis \cite{Bielefeld_Md}, that gives $m_{D}/T$ slowly 
decreasing with $T$  
($m_{D}/T\approx 3$ at $T\sim 1.5T_{c}$, $m_{D}/T\approx 2.4$ at 
$T\sim 4T_{c}$). 
As in our analyses \cite{RAA08,RAA13,RPP14} of JQ  
we use the one-loop running $\alpha_{s}$ 
frozen at low momenta at some value $\alpha_{s}^{fr}$. 
The analyses of the low-$x$ structure functions 
\cite{NZ_HERA} and 
of the heavy quark energy loss in vacuum \cite{DKT} show that for 
gluon emission in vacuum for this parametrization  
$\alpha_{s}^{fr}\approx 0.7-0.8$. But in the QGP the thermal effects 
can suppress the in-medium QCD coupling.
Our analysis of the LHC data on the nuclear modification factor 
$R_{AA}$ for Pb+Pb collisions at $\sqrt{s}=2.76$ TeV 
within the LCPI approach to the induced gluon emission gives
the value $\alpha_s^{fr}\approx 0.4$ \cite{RAA13}. We will use this value in
this work. 
In Fig.~6 we plot the
$\rho$-dependence of the dipole cross section for $gg$ state obtained
for several values of the QGP temperature. 
At small $\rho$ (say, $\lsim 0.1$ fm) the dipole cross section 
has nearly quadratic form 
$\sigma_{8}(\rho)\approx C\rho^2$, where $C$ depends logarithmically 
on $\rho$. In terms of the transport 
coefficient $\hat{q}$ \cite{BDMPS} and the effective number
density of the triplet color scattering centers 
one can write $C=\hat{q}/2n$. Our $\sigma_{8}(\rho)$ for $\rho\sim 0.1$ fm  
corresponds to $\hat{q}\sim 0.25$ at $T=250$ MeV. It agrees reasonably with
the qualitative pQCD calculations of 
Ref.~\cite{Baier_qhat} 
$
\hat{q}\sim 2\varepsilon^{3/4}\,
$
with $\varepsilon$ the QGP energy density (in terms the QGP temperature 
it is $\hat{q}\approx 15T^3$).

From Fig.~6 one sees that the dipole cross section flatten 
at $\rho\sim 2/m_D\sim 1/T$.
Fig.~6 shows that between the quadratic and flat regions
there is a rather broad region where 
the dipole cross section is approximately $\propto \rho$.
Note that for the regimes with the dipole cross section 
$\propto \rho^2$($\rho$),
for a given transverse momentum of the $gg$ pair and the 
longitudinal coordinate of 
the splitting point 
$z_s$, the probability of the off-diagonal transitions
falls approximately $\propto 1/E^{2N}$($1/E^{N}$), where  $N$ is the number 
of rescatterings. For this reason, as will be seen below,
the production of the decuplet $gg$ states, that requires
$N\ge 2$, falls steeply with the jet energy.

\section{Numerical results}
We perform analysis of the $z$-dependence of the probability
distribution vector $\vec{P}$ for the $gg$ state averaged over the 
internal momentum of the $gg$ system.
We describe the gluon $(x,\qb)$-distribution for the $g\to gg$ transition
by the leading order pQCD formula
\beq
\frac{dN}{dxd\qb}=\frac{C_{A}\alpha_{s}(q^2)}{\pi^{2}}\Big[
\frac{1-x}{x}+\frac{x}{1-x}+x(1-x)\Big]\frac{q^{2}}{(q^{2}+\epsilon^{2})^{2}}\,,
\label{eq:350}
\eeq
where $C_A$ is the gluon color Casimir factor, and
$\epsilon^{2}=m_{g}^{2}(1-x+x^{2})$. Here the effective gluon mass $m_g$
plays the role of the infrared cutoff. 
For numerical computations 
we take  $m_{g}=0.75$ GeV. This value was obtained  
from the analysis 
of the low-$x$ proton structure function $F_{2}$
within the dipole BFKL equation
\cite{NZ_HERA}. It agrees well with the natural infrared cutoff 
for perturbative gluons $m_{g}\sim 1/R_{c}$, where 
$R_{c}\approx 0.27$ fm is the gluon correlation radius in the QCD vacuum
\cite{shuryak1}. 
For a given $x$ we restrict the value of $q$ 
by $q_{max}=Ex(1-x)$, where $E$ is the energy of the parent gluon.
In calculation of the distribution (\ref{eq:350}) we use $\alpha_s$ frozen
at the value $\alpha_{s}^{fr}=0.7$. This value was previously obtained 
by fitting the data on proton structure function
$F_{2}$ at low $x$
within the dipole BFKL equation \cite{NZ_HERA}. 
This value is also consistent with the relation 
\beq
\int_{\mbox{\small 0}}^{\mbox{\small 2 GeV}}\!dQ\frac{\alpha_{s}(Q^{2})}{\pi}
\approx 0.36 \,\,
\mbox{GeV}\,
\label{eq:360}   
\eeq
obtained in \cite{DKT} from the analysis of the heavy quark energy 
loss in vacuum.
\begin{figure}
\hspace*{-0.8cm }\includegraphics[height=5cm]{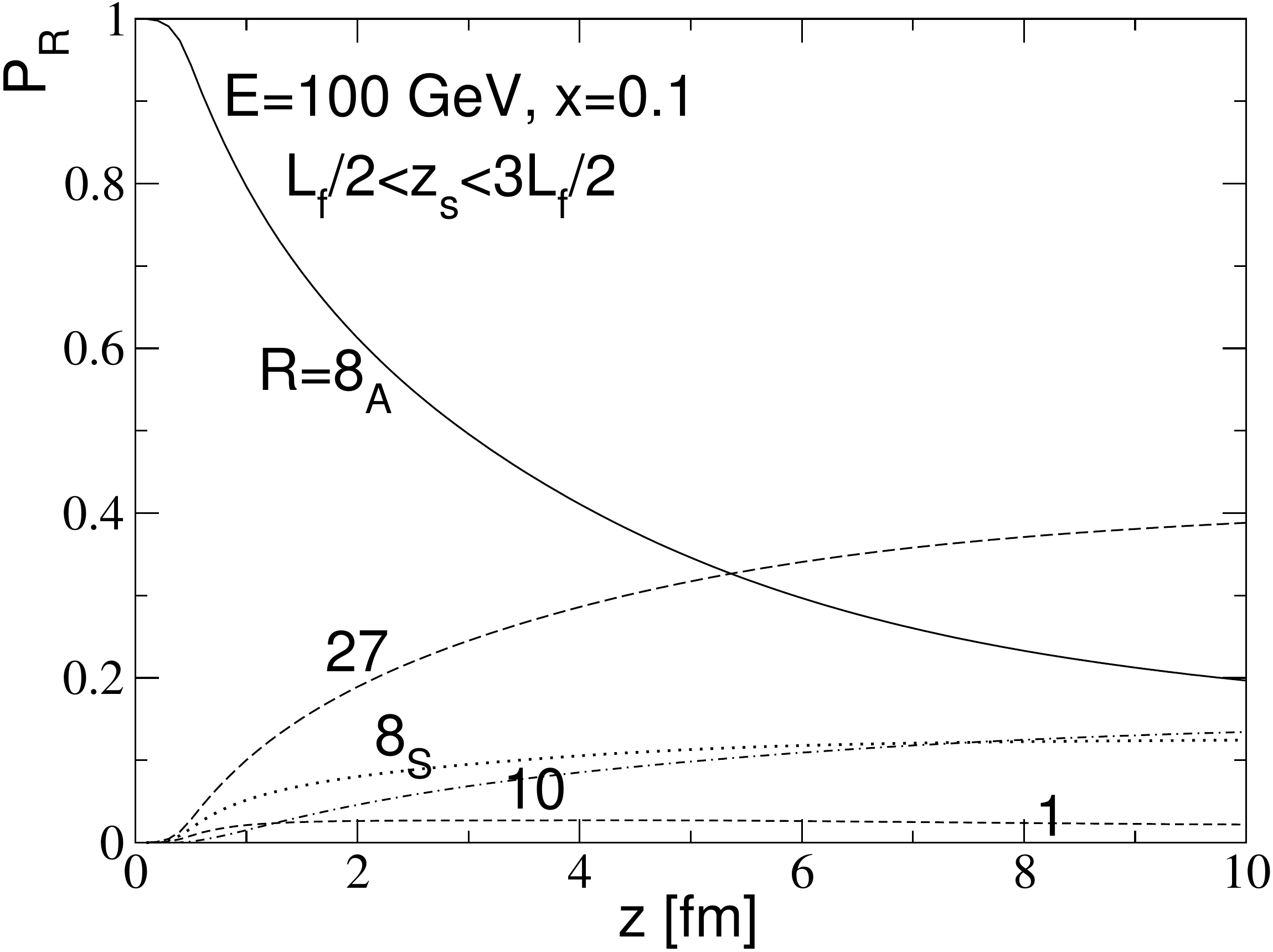}
\hspace*{0.8cm }\includegraphics[height=5cm]{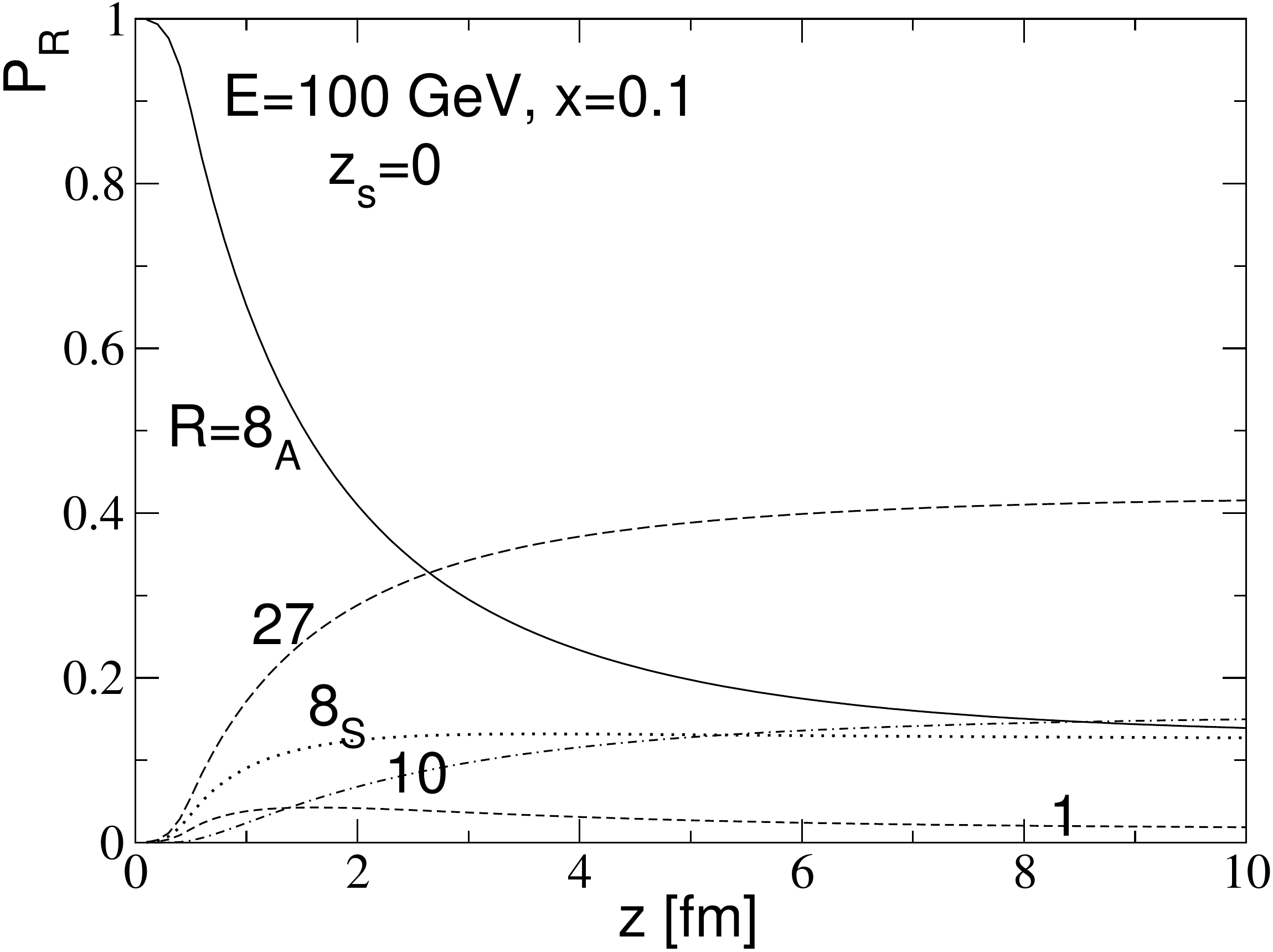}
\caption{\small 
Coefficients $P_R$ for $g\to gg$ splitting with $x=0.1$ averaged over
the transverse momentum versus the jet path length for the initial
gluon energy $E=100$ GeV (see main text for details). 
Left: The position of the splitting point 
in the interval $L_f/2<z_s<3L_f/2$. Right: The splitting
at $z_s=0$. Dashed: $R=1$; Solid: $R=8_A$; Dotted: $R=8_S$; Long-dashed:
$R=27$; Dot-dashed: $R=10$.
}
\end{figure}
\begin{figure}
\hspace*{-0.8cm }\includegraphics[height=5cm]{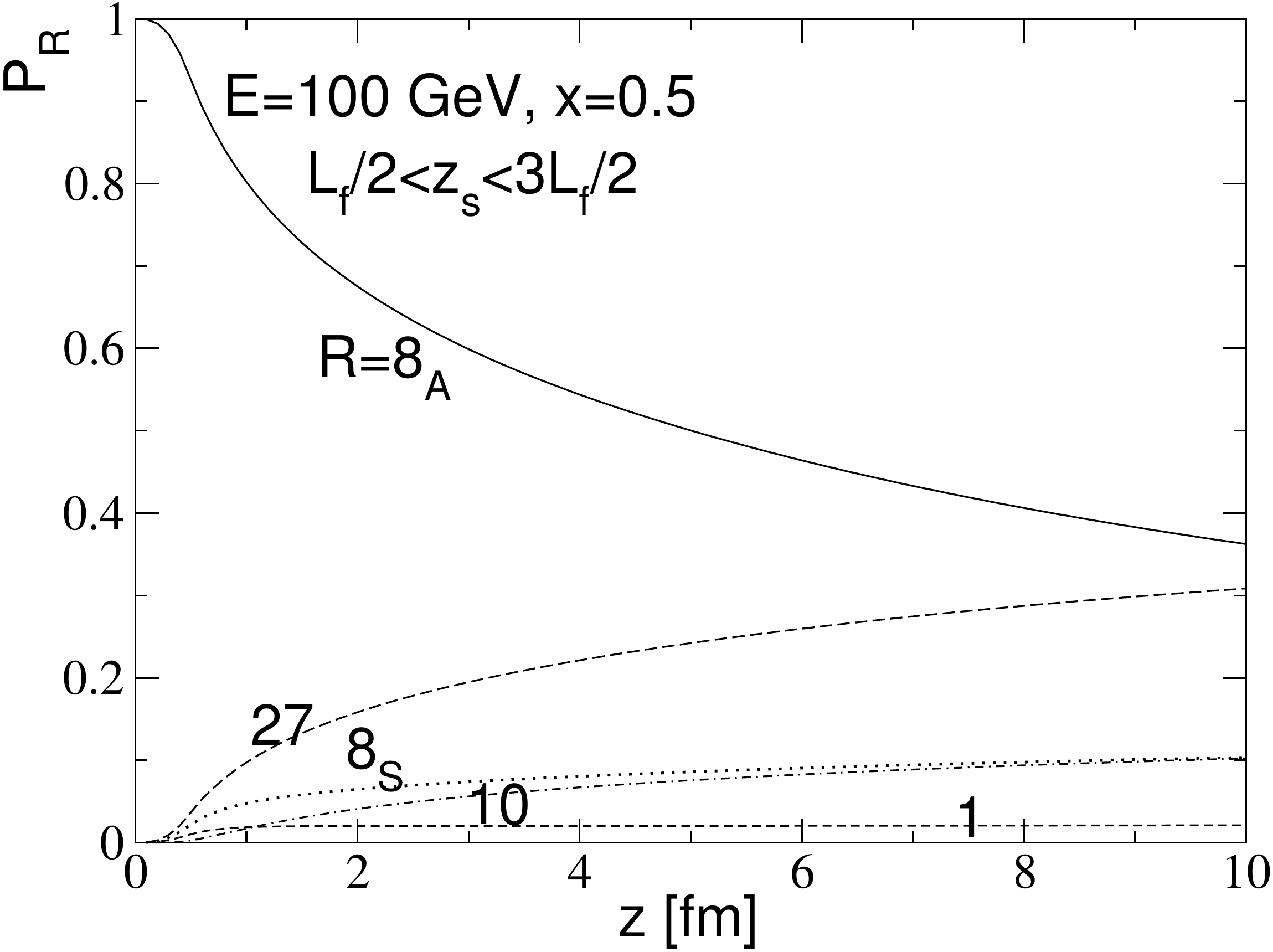}
\hspace*{0.8cm }\includegraphics[height=5cm]{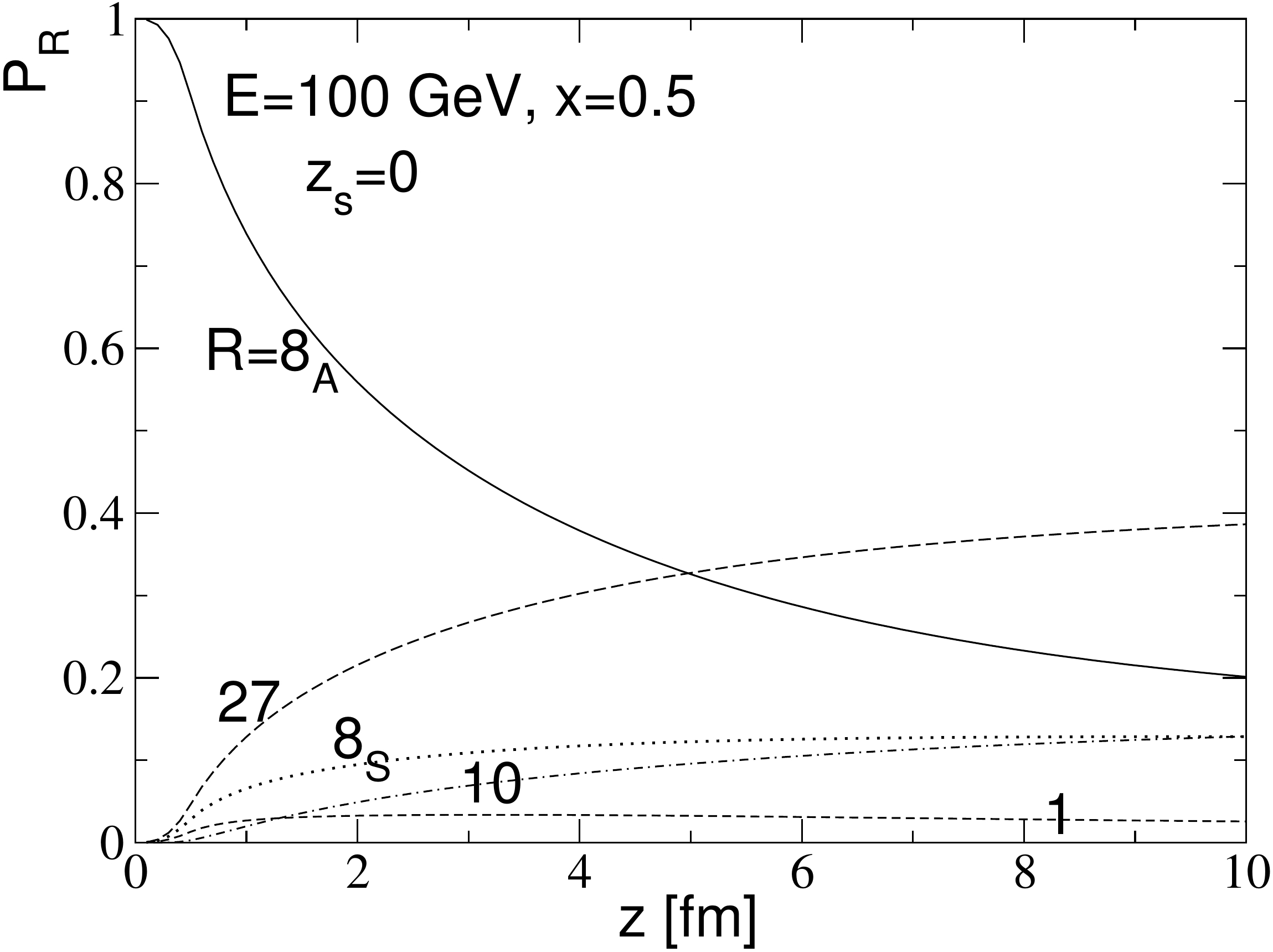}
\caption{\small 
Same as in Fig.~7, $E=100$ GeV, $x=0.5$.
}
\end{figure}
\begin{figure}
\hspace*{-0.8cm }\includegraphics[height=5cm]{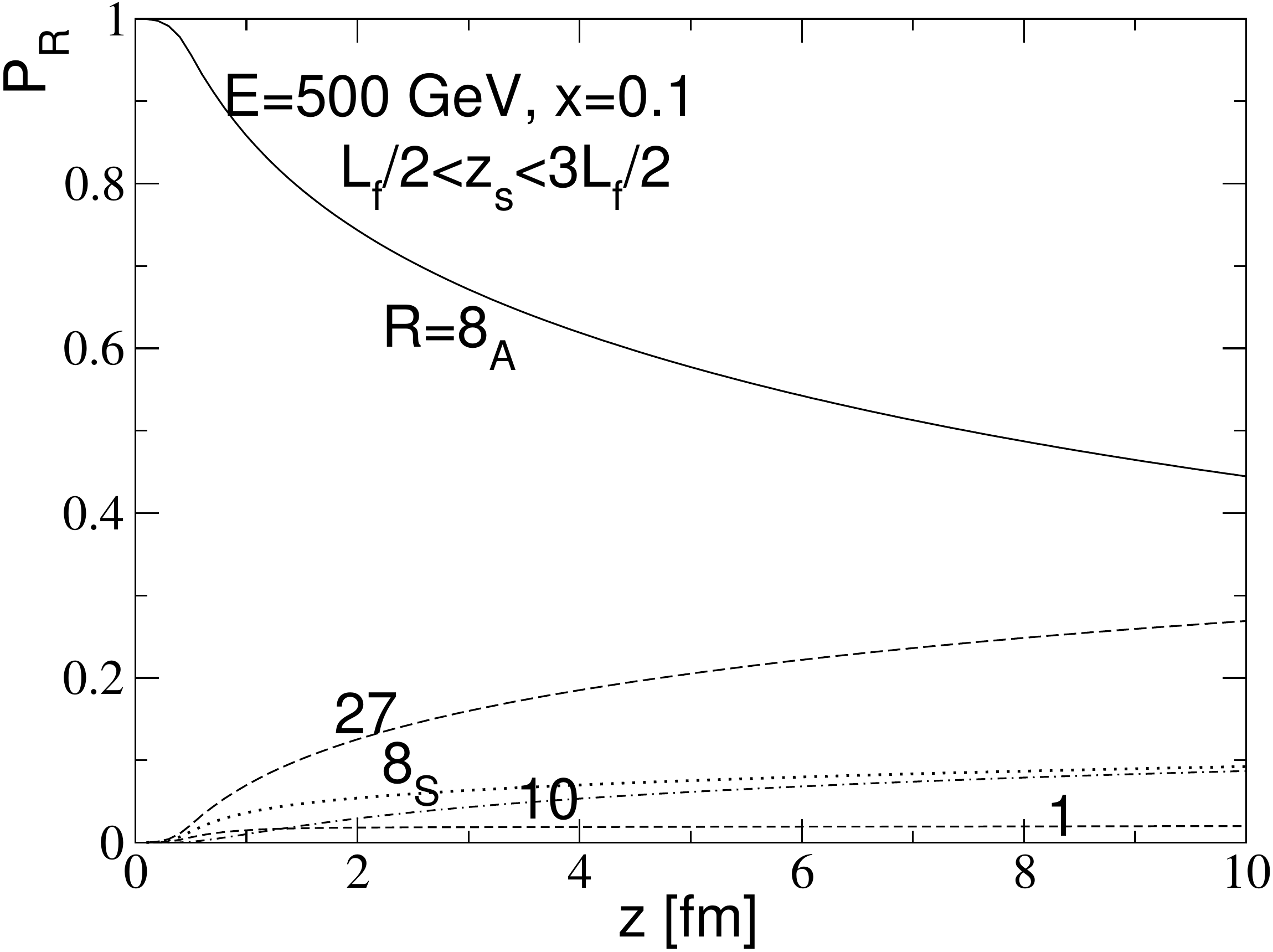}
\hspace*{0.8cm }\includegraphics[height=5cm]{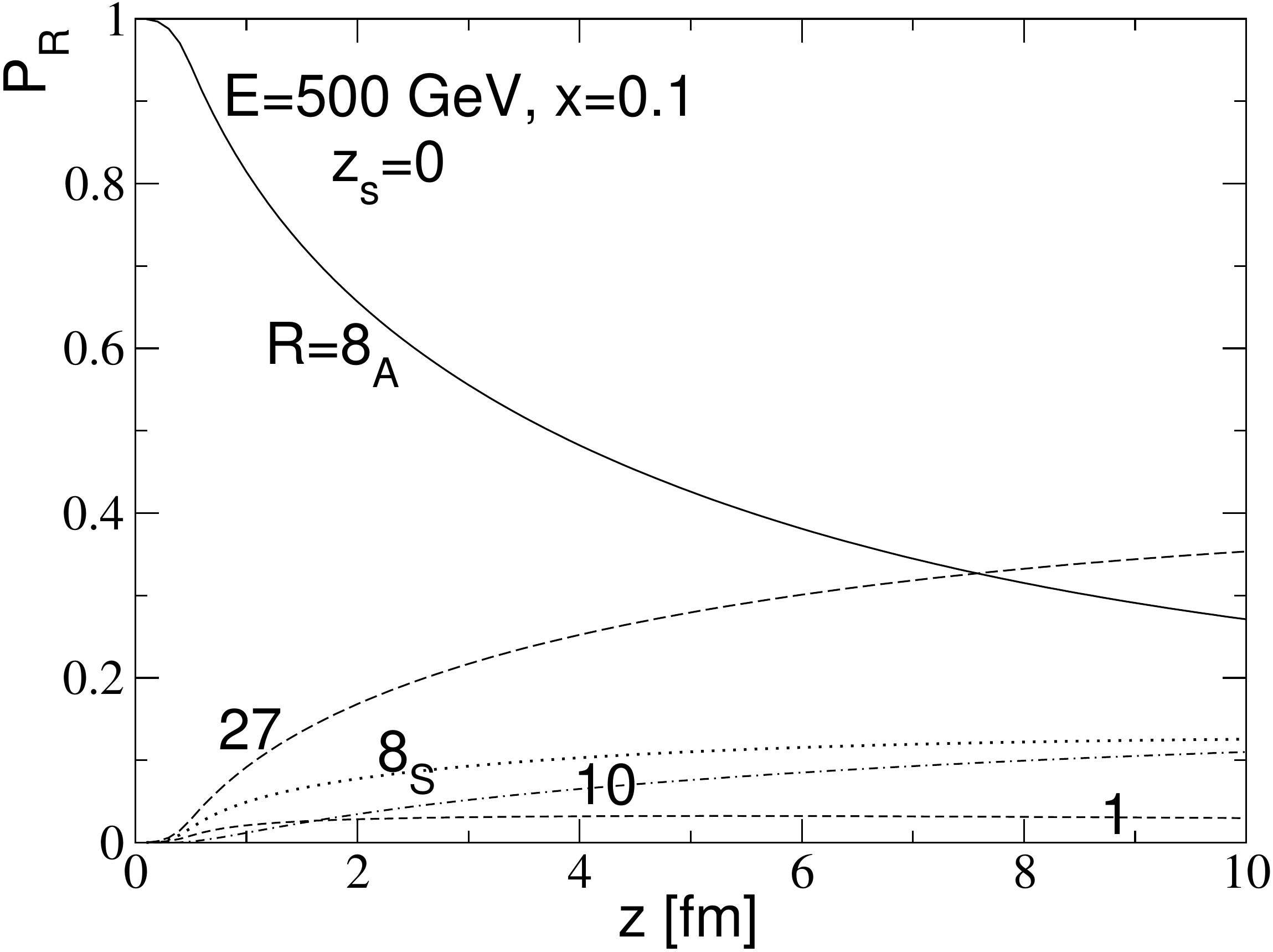}
\caption{\small 
Same as in Fig.~7, $E=500$ GeV, $x=0.1$.
}
\end{figure}
\begin{figure}
\hspace*{-0.8cm }\includegraphics[height=5cm]{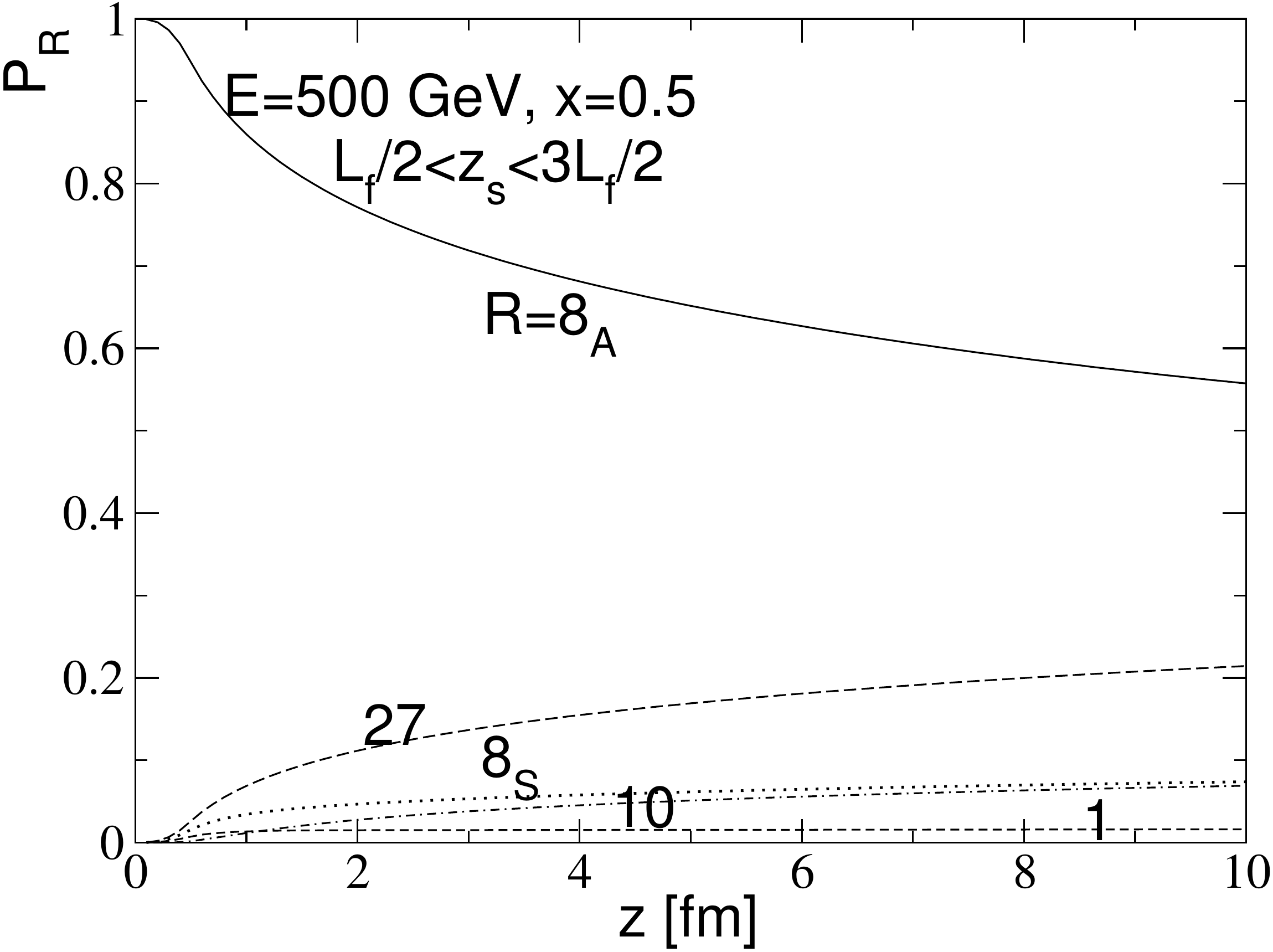}
\hspace*{0.8cm }\includegraphics[height=5cm]{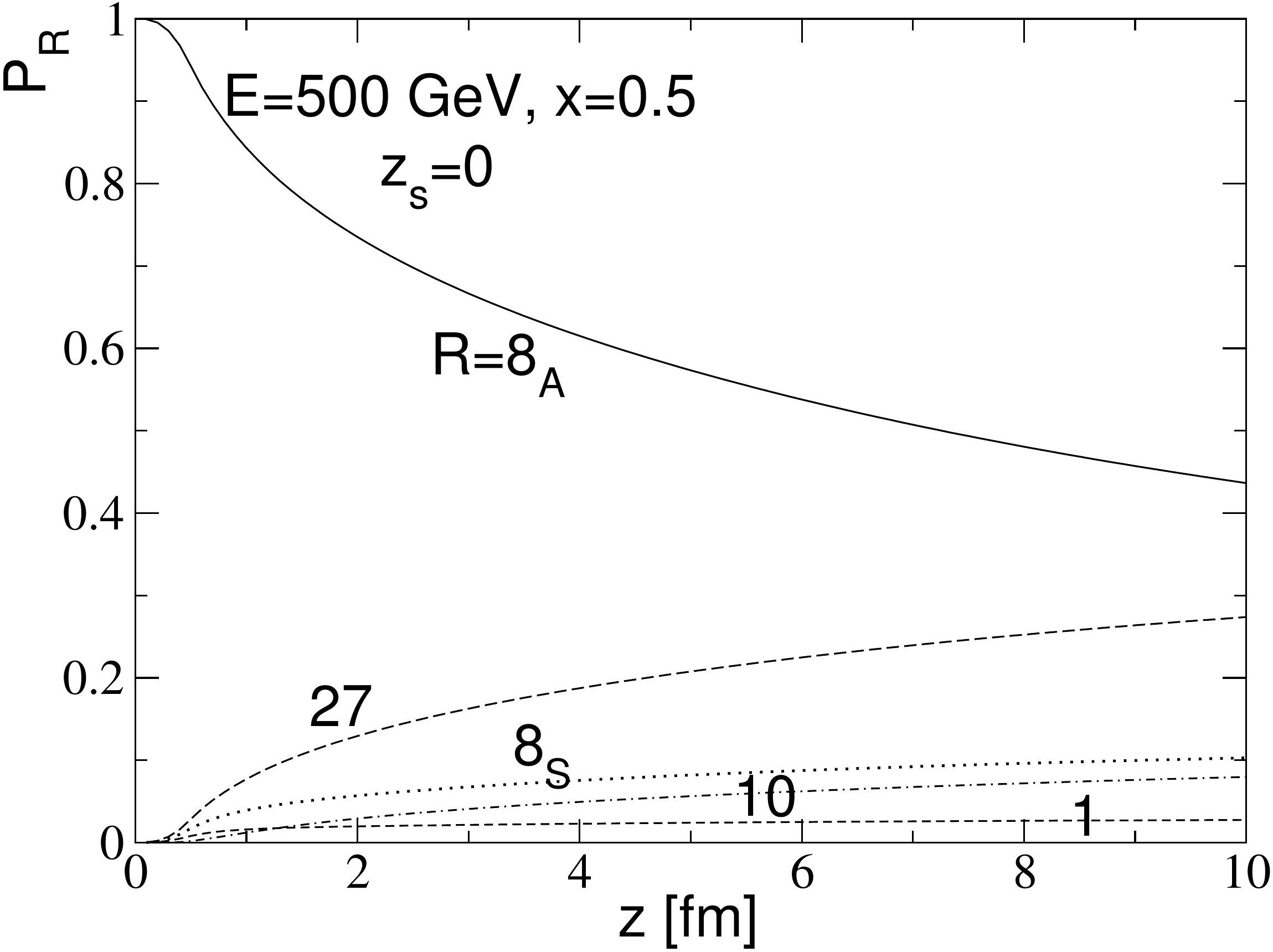}
\caption{\small 
Same as in Fig.~7, $E=500$ GeV,  $x=0.5$.
}
\end{figure}

We define the $q$-averaged $P_R$ as
\beq
P_R(z)=\frac{\int d\qb P_R(z,q)
\frac{dN}{dxd\qb}}{\int d\qb 
\frac{dN}{dxd\qb}}\,,
\label{eq:370}   
\eeq
where $P_R(z,q)$ is the solution to the evolution equation
(\ref{eq:200}) for a given geometry of the $g\to gg$ splitting.
The transverse momentum $q$ defines the angle between gluon momenta of 
the final $gg$ pair. The geometry of the gluon trajectories also depends
on the longitudinal coordinate $z_s$ of the splitting point.
From the uncertainty relation $\Delta p_z\Delta z\sim 1$
one can obtain for the typical formation length of the $gg$ 
pair
\beq
L_f\sim \frac{2x(1-x)E}{q^2+\epsilon^2}\,.
\label{eq:380}   
\eeq
We perform computations for two versions: for instant
decay of the primary gluon, i.e., $z_s=0$, and delayed decay,
when the splitting point is uniformly distributed 
in the interval  $L_f/2<z_s<3L_f/2$.
In Figs.~7--10 we present the results for $P_R(z)$ 
obtained for $E=100$ and $500$ GeV
for $g\to gg$ splitting with $x=0.1$ and $0.5$.
As one can see, the color randomization becomes stronger
with decreasing fractional longitudinal momentum $x$. It is due to 
growth of the angle between gluons
in the asymmetric $gg$ pairs.  
For the version with the delayed  decay  the
color randomization is noticeably weaker than that for the
instant decay. However, even in the latter case the color randomization
of the two gluon system turns out to be rather slow. 
For example, one can see that for the typical
jet path length $L\sim 5$ fm for central Pb+Pb collisions
the probability for the $gg$ pair to stay in the $8_A$ state
differs substantially from that in the regime of the full color randomization
(especially for symmetric splitting).
From Figs.~7--10 one sees that the color randomization is weakest for the
decuplet states. This occurs because, contrary to the multiplets 
$1$, $8_S$, $27$,  the direct $N=1$ rescattering transition 
of the $8_A$ state to the decuplet states is forbidden.  And the leading
order in the QGP density $N=2$ rescattering contribution comes
from the sequential transitions: $8_A\to 8_S,27\to 10(\overline{10})$.

\begin{figure}
\hspace*{-0.8cm }\includegraphics[height=5cm]{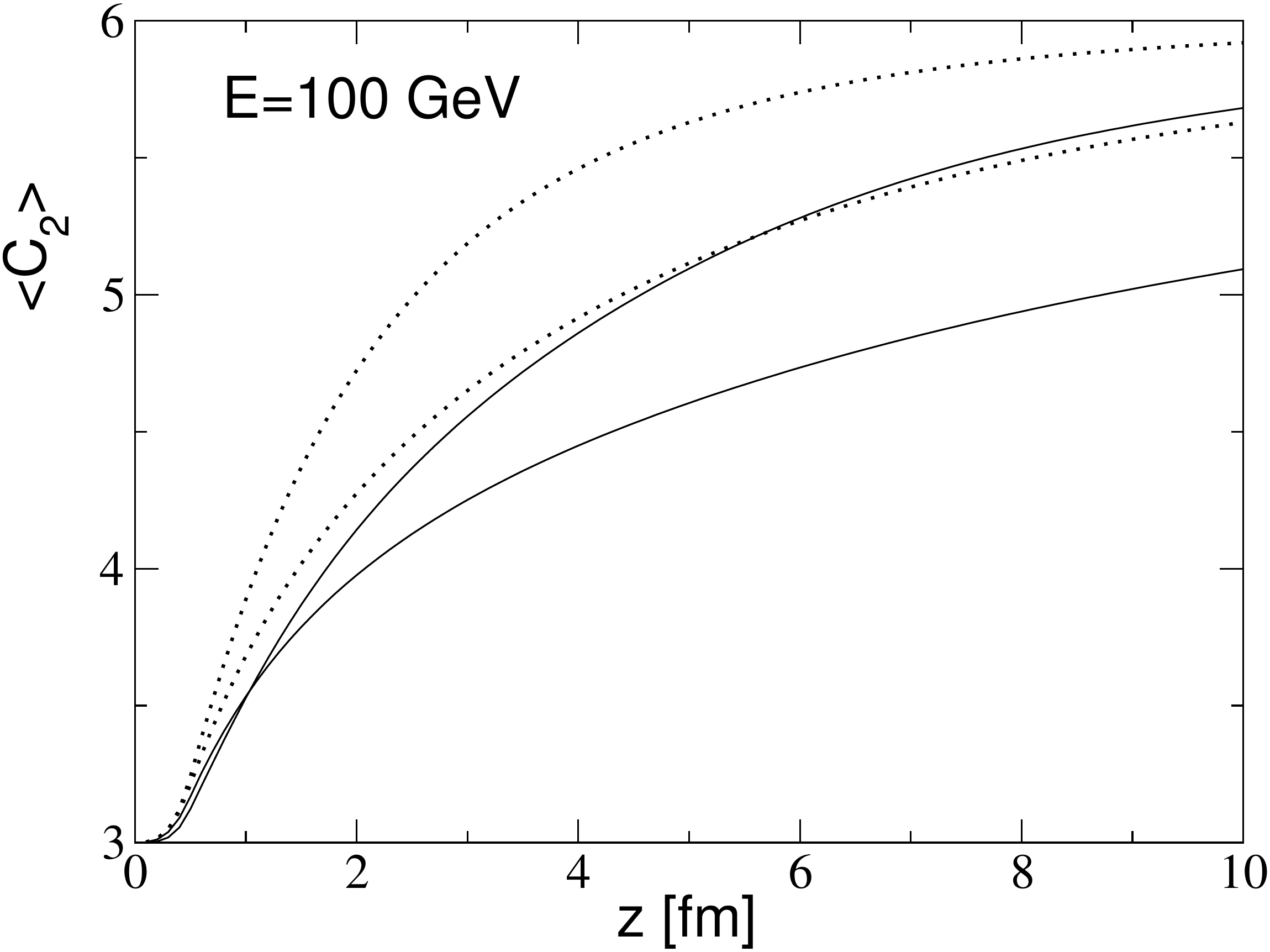}
\hspace*{0.8cm }\includegraphics[height=5cm]{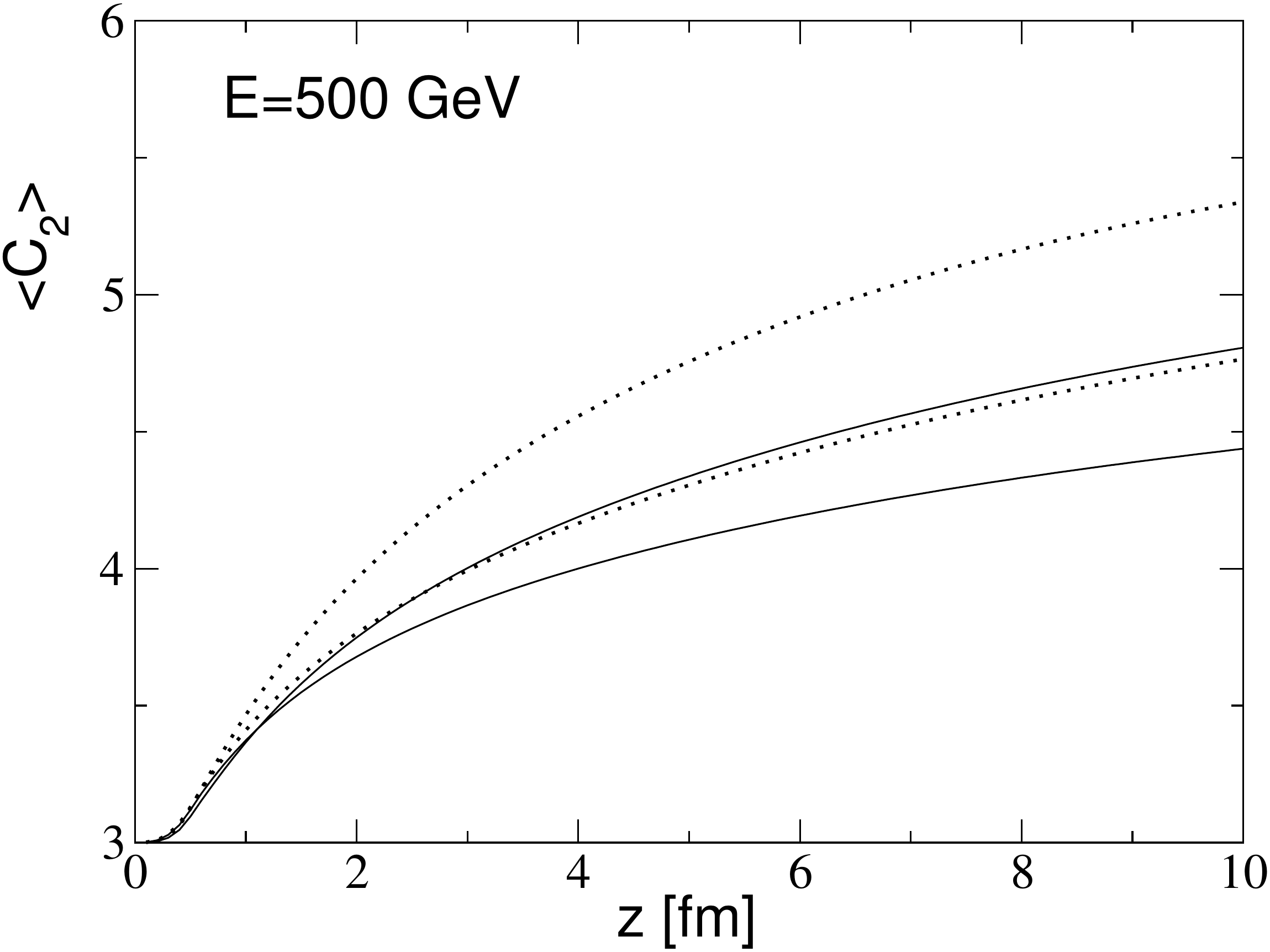}
\caption{\small 
The multiplet-averaged Casimir operator 
$\langle C_2\rangle$
for $g\to gg$ splitting
with $x=0.1$ and $x=0.5$ (from top to bottom) 
at $E=100$ GeV (left) and $E=500$ GeV (right) versus the jet path length
obtained for the splitting point 
in the interval $L_f/2<z_s<3L_f/2$ (solid) and for the splitting
at $z_s=0$ (dashed).
}
\end{figure}
\begin{figure}
\includegraphics[height=5cm]{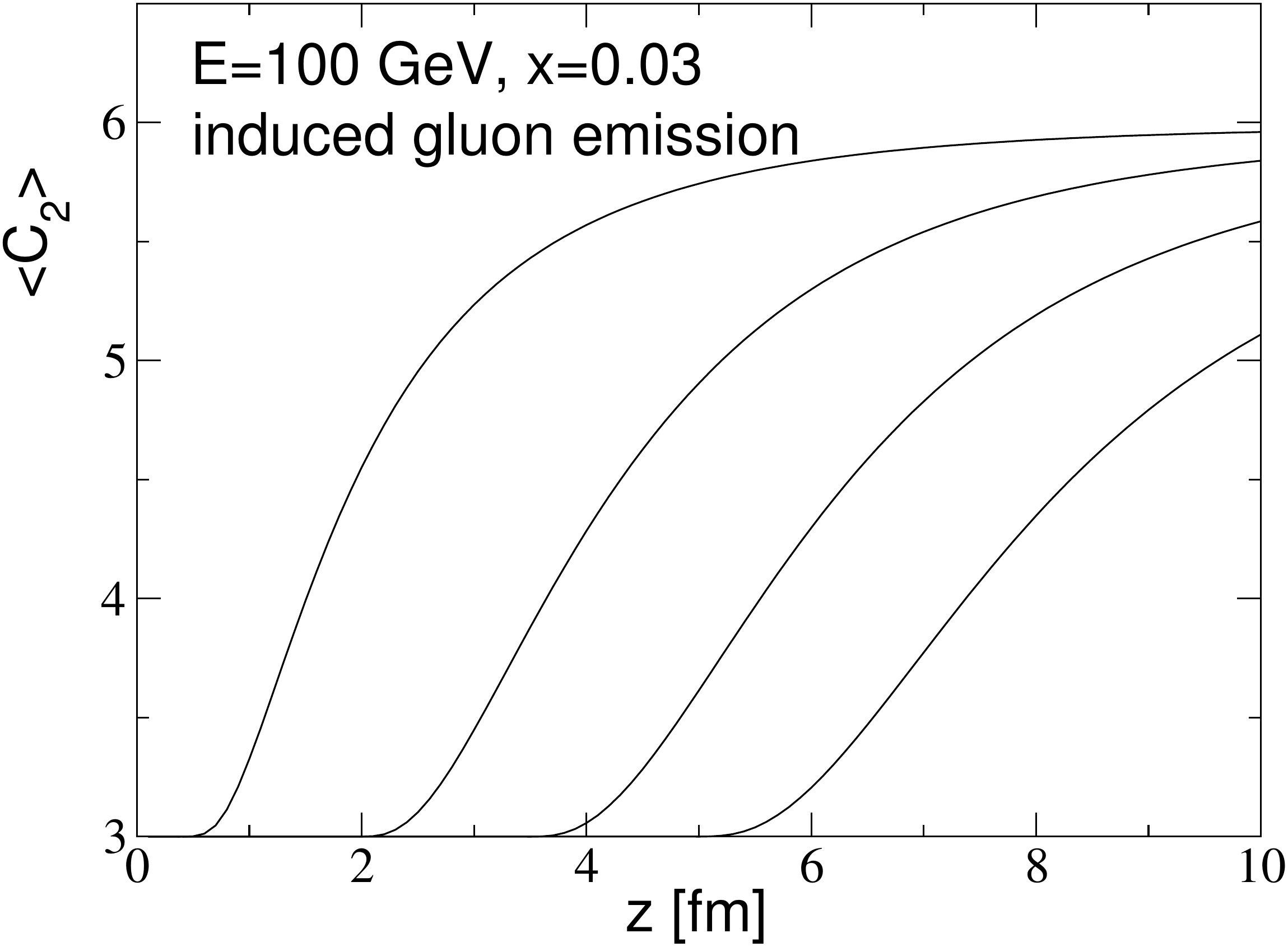}
\caption{\small 
The multiplet-averaged Casimir operator for soft induced 
$g\to gg$ splitting with $x=0.03$
at $E=100$ GeV versus the jet path length
obtained for the splitting point 
(from left to right)
$z_s=0.5$, $2$, $3.5$ and $5$ fm
(see main text for details).
}
\end{figure}
\begin{figure}
\includegraphics[height=5cm]{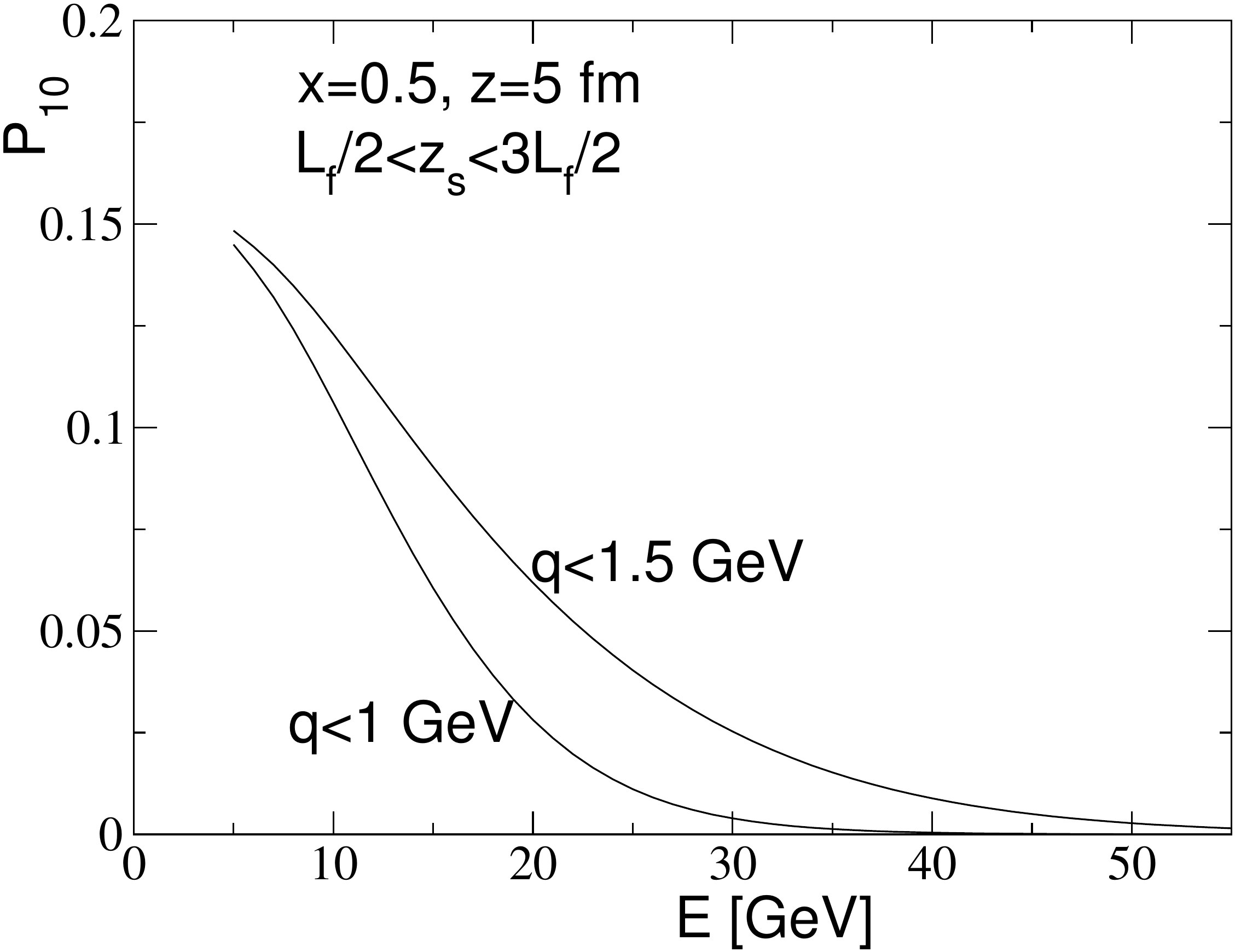}
\caption{\small 
Energy dependence of the probability
of the decuplet state  for $q<1$ and $q<1.5$ GeV for the typical
jet path length in the QGP $z=5$ fm. The splitting $g\to gg$ occurs 
in the region   $L_f/2<z_s<3L_f/2$.
}
\end{figure}

Besides the probabilities for distinct multiplets it is instructive to 
examine the $z$-dependence of the 
$SU(3)$-multiplet averaged color Casimir of the $gg$ pair
\beq
\langle C_2\rangle=\sum_R P_RC_2[R]\,,
\label{eq:390}   
\eeq
that may be viewed as a reasonable integral characteristic of the 
color randomization. 
Instantly after $g\to gg$ splitting 
$\langle C_2\rangle=C_2[8]=N_c$, and for fully randomized regime   
$\langle C_2\rangle=2C_2[8]=2N_c$.
As was mentioned above, the total Casimir of the $gg$
pair controls the emissions of gluons with the inverse transverse
momentum smaller than the transverse size of the $gg$ pair, when it 
acts as a single radiator. 
In Fig.~11 we plot the $z$-dependence of $\langle C_2\rangle$. 
As in the figures for $P_R$, we show the results for the versions with 
the instant and delayed splitting.
From Fig.~11 one sees that $\langle C_2\rangle$ for the typical jet path length 
$L\sim 5$ fm for central Pb+Pb collisions differs considerably from
its value $2N_c$ for the fully color randomized $gg$ state.
At $E=500$ GeV even at $L=10$ fm 
the color randomization is not reached. In this case 
$\langle C_2\rangle$  lies in the middle between 
the values for the pure octet state and for the fully 
randomized two-gluon state (both for the asymmetric ($x=0.1$) 
and symmetric ($x=0.5$) gluon pairs).

As we said, the approximation of rigid geometry 
with the straight-line gluon trajectories (the same for 
amplitude and for the complex conjugate one)
seems to be reasonable for a hard $g\to gg$
splitting.
For the soft induced gluon emission the quantum fluctuations
of the parton trajectories are important, and the full path
integral machinery should be used.
In this case the  color algebra part of the calculations
should be performed before the integrations over the parton trajectories.
And the effect of the randomization in the color space and the effect of the
fluctuations of the trajectories cannot be separated.
Nevertheless, the approximation of rigid
geometry seems to be a reasonable method for a qualitative analysis
of the color randomization for the induced $g\to gg$ splitting. 
To perform such an analysis we use estimates of the local formation length
and the transverse momentum squared for 
induced gluon emission inside the QGP 
in terms  of the transport coefficient $\hat{q}$
(see, e.g. \cite{BDMPS,Mueller_model}): $k_T^2\sim \sqrt{2\omega \hat{q}}$,
and $L_f^{in}\sim \sqrt{2\omega/\hat{q}}$. We take the local transport
coefficient in the form
$\hat{q}\approx 2\varepsilon^{3/4}\approx 15T^3$ obtained in \cite{Baier_qhat},
which, as we said above, is in a reasonable agreement with 
our dipole cross section $\sigma_8(\rho)$ at small $\rho$. 
For the induced $g\to gg$ splitting we use the Gaussian transverse 
momentum distribution ${dN}/{dk^2}\propto \exp{(-k_T^2/\langle k_T^2\rangle)}$
with  $\langle k_T^2\rangle=\sqrt{2\omega \hat{q}}$, and the local $\hat{q}$
calculated at $L=z_s+L_f^{in}/2$. 
In Fig.~12 we present the average Casimir operator for 
the soft $g\to  gg$ splitting  at $E=100$ GeV for $x=0.03$ 
(i.e. at $\omega=3$ GeV) obtained in this
model for the splitting positions in the QGP $z_s=0.5$, $2$, $3.5$ and $5$ fm.
Note that at $x\ll 1$ the results are insensitive to the value of $E$.
From Fig.~12 one can see that even for the induced gluon emission
in the initial hottest stage of the QGP the color randomization
requires $\sim 5$ fm. This plot clearly shows that for gluons emitted at 
later stages ($L\sim 3-5$ fm) the color randomization is incomplete even
at $L\sim 10$ fm. Thus, our results show that a considerable part
of the gluon pairs created in the induced $g\to gg$ processes may leave 
the QGP without complete color decoherence.

The formalism of the present paper allows to study quantitatively 
the energy dependence of the decuplet $gg$ pair production. As we said
in introduction, it is important for better understanding of the role
of the collinear decuplet two-gluon states in the baryon jet fragmentation
due to the mechanism shown in Fig.~2.
In Ref.~\cite{AZ_baryon} the contribution of this mechanism has been estimated
in the approximation of complete color randomization.
The analysis of  Ref. \cite{AZ_baryon} is based on the calculation of 
the phase space for formation of the diquark (that is formed from two quarks
created after the conversion of both gluons to $q\bar{q}$ pairs,
as shown in Fig.~2) with mass $M_D\lsim 1-1.5$ GeV. 
The diquark configurations with higher values of $M_D$ 
should have a smaller probability for fragmentation into
a leading baryon. The dominating contribution to such diquark states
comes from the gluon splitting to symmetric gluon pairs with $x\sim 0.5$
and $q\lsim 1$ GeV.
The qualitative estimates  of Ref.~\cite{AZ_baryon} show
that the approximation of the complete color randomization
of the $gg$ pairs should stay reasonable for jets with $E\lsim 20-30$ GeV.
The corresponding limit for the baryon momenta is smaller by a factor 
of $\sim 2-3$ (because the diquark momentum is smaller than the jet energy 
by a factor of $\sim 2$, and some part of the longitudinal momentum is lost
in the diquark fragmentation to the observed baryon). 
This means that the contribution
of the anomalous color decuplet $gg$ pairs may be potentially important
for baryon production at $p_T\lsim 7-10$ GeV. Beyond 
this $p_T$ region it should decrease steeply due to the fall of the probability
to find the $gg$ pair in the decuplet color state.
Because for a given $q$ the angle  between gluons 
 $\propto 1/E$ and the transverse size of the $gg$ 
pair also decreases $\propto 1/E$. As a result, the probability 
of excitation of the decuplets states
should fall steeply with $E$, as was already said in Sec.~3.  
However, of course, the estimates of Ref.~\cite{AZ_baryon} are very crude.
We use our formalism to perform a quantitative analysis.
In Fig.~13 we plot the energy dependence of the probability
of the decuplet state  for $q<1$ and $q<1.5$ GeV for 
$L=5$ fm, which is the typical jet path length in the QGP for central
Pb+Pb collisions.
The curves are obtained 
for the splitting $g\to gg$ in the region  $L_f/2<z_s<3L_f/2$.
From Fig.~13 one sees that for jets with $E\sim 5-7$ GeV
$P_{10}$ is close to that for full color randomization, i.e. $P_{10}=10/64$,
and from $E\sim 10$ GeV to  $E\sim 30$ GeV $P_{10}$ falls steeply.
In terms of the baryon transverse momentum it means
the anomalous contribution of the baryon production
becomes small at $p_T\gsim 10-15$ GeV. 
This is in agreement with the recent data from ALICE \cite{ALICE_baryon}
on the high-$p_T$ spectra in Pb+Pb collisions
at $\sqrt{s}=2.76$ TeV that show that the ratio 
$(p+\bar{p})/(\pi^++\pi^-)$ becomes close to that for $pp$ collisions
at $p_T\gsim 10-15$ GeV. 
The steep decrease of the probability for production
of the decuplet $gg$ states is a consequence of the fact that 
the excitation of the decuplet state requires, as was said before, 
at least two rescatterings
and growth of the formation length (that leads to reduction of the 
effective path length of the $gg$ pair in the QGP).

\section{Conclusions}

In this paper we have studied
the dynamics of the color randomization of two-gluon states 
produced after splitting of a primary fast gluon in
the QGP formed  in heavy ion collisions.  Numerical calculations have
been performed for central Pb+Pb collisions at the LHC energy 
$\sqrt{s}=2.76$ TeV.
The analysis is based on the evolution equation for the color density
matrix for $gg$ system obtained in the dipole approach. In our framework 
the density matrix of the two-gluon pair may be viewed as 
wave function of a color singlet four-gluon system.
The $L$-dependence of this wave function is controlled by the diffraction
operator for scattering of the four-gluon system on the QGP constituents.
We have found that the color randomization of the
$gg$ pairs turns out to be rather slow. 
Our calculations show that for jet energies  $E=100$ and $500$ GeV 
the $SU(3)$-multiplet averaged color Casimir $C_2$ of the $gg$ state 
for the typical jet path length 
$L\sim 5$ fm for central Pb+Pb collisions differs considerably from
its value $2N_c$ for the fully color randomized $gg$ state.
For jets with $E=500$ GeV even at $L=10$ fm the color randomization 
is not reached. In this case the averaged color Casimir lies in 
the middle between the values for the pure octet state (as for the 
parent gluon) and for the fully 
randomized two-gluon state (both for the asymmetric ($x=0.1$) 
and symmetric ($x=0.5$) gluon pairs).

We have found that the rate of the color randomization
is slowest for the decuplet color multiplets: $10$ and $\overline{10}$.
This occurs because, contrary to other states in the Clebsch-Gordan 
decomposition of the direct product of two octets,
 the direct transition $8_A\to 10(\overline{10})$ for $N=1$ rescattering
is forbidden, and the excitation of the decuplet states
goes through  excitation of the intermediate $8_S$ and $27$ multiplets.

We have studied the energy dependence of the generation of the 
nearly collinear decuplet $gg$ states, that can lead to production
of leading baryons in jet fragmentation \cite{AZ_baryon,M98}.
We find that the probability to observe such pairs decreases steeply 
with jet energy, and becomes very small for $E\gsim 30$ GeV.
It allows one to conclude that the contribution
of this mechanism to the baryon production should become very small 
at $p_T\sim 10$ GeV. This agrees reasonably with the data from ALICE
\cite{ALICE_baryon} on the ratio
$(p+\bar{p})/(\pi^++\pi^-)$
in Pb+Pb collisions at $\sqrt{s}=2.76$ TeV.

\begin{acknowledgments} 	
This work has been supported by the RScF grant 16-12-10151.
\end{acknowledgments}

\section*{Appendix}

The contribution to the two-gluon color wave function 
$\langle ab |\Psi\rangle$ 
of a given irreducible $SU(3)$ multiplet
$R$ in the Clebsch-Gordan decomposition (\ref{eq:40}) can be written as 
$P[R]^{ab}_{cd}\langle cd |\Psi\rangle$, where $P[R]$ is the projector
operator for the multiplet $R$ given by 
the general quantum mechanical formula (\ref{eq:140}). 
The projectors onto the multiplets
$1$, ${8_A}$, $8_S$, $27$, $10$ and $\overline{10}$  can be written in terms
of the Kronecker deltas, antisymmetric
tensor $f_{abc}$ and symmetric tensor $d_{abc}$ as
\beq
P[1]^{ab}_{cd}=\frac{1}{8}\delta_{ab}\delta_{cd}\,,
\label{eq:400}   
\eeq  
\beq
P[8_A]^{ab}_{cd}=\frac{1}{3}f_{abk}f_{kcd}\,,
\label{eq:410}   
\eeq  
\beq
P[8_S]^{ab}_{cd}=\frac{3}{5}d_{abk}d_{kcd}\,,
\label{eq:420}   
\eeq  
\beq
P[27]^{ab}_{cd}=\frac{1}{2}(\delta_{ac}\delta_{bd} +
\delta_{ad}\delta_{bc})-\frac{1}{8}\delta_{ab}\delta_{cd}  
-\frac{3}{5}d_{abk}d_{kcd}\,,
\label{eq:430}   
\eeq  
\beq
P[10]^{ab}_{cd}=\frac{1}{4}(\delta_{ac}\delta_{bd} -
\delta_{ad}\delta_{bc})-
\frac{1}{6}f_{abk}f_{kcd}+\frac{i}{2}Y^{ab}_{cd}\,,
\label{eq:440}   
\eeq
\beq  
P[\overline{10}]^{ab}_{cd}=\frac{1}{4}(\delta_{ac}\delta_{bd} -
\delta_{ad}\delta_{bc})-
\frac{1}{6}f_{abk}f_{kcd}-\frac{i}{2}Y^{ab}_{cd}\,,
\label{eq:450}   
\eeq  
where
\beq
Y^{ab}_{cd}=-\frac{1}{2}(d_{ack}f_{kbd}+  f_{ack}d_{kbd})\,.
\label{eq:460}   
\eeq

The two-gluon color wave function
for the states $1$, $8_S$, $27$ are 
symmetric in permutations of gluon color indexes
$a\leftrightarrow b$ and $c\leftrightarrow d$, and the 
states $8_A$, $10$, $\overline{10}$ are antisymmetric.
The contribution of the first three terms in the formulas for the decuplet
projectors is clearly antisymmetric, the fact that the term $Y^{bc}_{cd}$ 
is also antisymmetric for $a \leftrightarrow b$ and
$c \leftrightarrow d$ is evident from the identity
\beq
  d_{ack}f_{kbd} + f_{ack} d_{kbd} =  d_{cbk}f_{kda} + f_{cbk} d_{kda} \,.
\label{eq:470}
\eeq
Calculations of the projectors for 
$1$, $8_S$, $27$ and $8_A$ multiplets are trivial, 
but for $10$ and $\overline{10}$ multiplets they are more complicated.
The projectors $P[10]$ and $P[\overline{10}]$ can be obtained
after straightforward (somewhat tedious) calculations with the help of 
the standard formula (\ref{eq:140}) using for the decuplet (antidecuplet) states
the symmetric spinor tensors $\Psi^{ijk}$ ($\Psi_{ijk}$).
For the two-gluon state these tensors can be built using
the spinor form of the gluon wave function 
$(g_a)^i_k=\frac{1}{\sqrt{2}}(\lambda_a)^i_k$.

An important fact for our calculations is that 
the projectors are proportional to the four-gluon color wave functions of the color singlets $|R\bar{R}\rangle$
built from $R$ and $\bar{R}$ multiplets (\ref{eq:160}).
The fact that the wave function
given by (\ref{eq:160}) describes a color singlet can be checked by calculating 
the expectation value 
\beq
\langle R\bar{R}|T^{\alpha}|R\bar{R}\rangle
\label{eq:480}   
\eeq
of the total color generator
$T^{\alpha}=\sum_{i=1}^4T^{\alpha}_i$ for four gluons, which should vanish
for color singlet states. One can easily show that this is true. 
Indeed, say, for the contribution from $i=1$ we have
\beq
\langle R\bar{R}|T^{\alpha}_1|R\bar{R}\rangle\propto
(P[R]^{a'b}_{cd})^*f_{\alpha a'a}P[R]^{ab}_{cd}\,.
\label{eq:490}   
\eeq
Since $(P[R]^{a'b}_{cd})^*=P[R]_{a'b}^{cd}$ and 
$P[R]^{ab}_{cd}P[R]_{a'b}^{cd})=P[R]^{ab}_{a'b}\propto \delta_{aa'}$
one can see that the left-hand side of (\ref{eq:490}) 
$\propto f_{\alpha aa}=0$. The fact that the projector operator
(\ref{eq:140}) is proportional to the color singlet wave function 
$|R\bar{R}\rangle$ is not surprising, because the last factor
on the right-hand side of (\ref{eq:140}) is proportional
to the wave function of the complex conjugate state 
$\langle cd |\bar{R}\bar{\nu}\rangle$ (where the component $\bar{\nu}$
has the ``magnetic'' quantum numbers opposite to that for $\nu$)
with the phase factor similar to that in the Clebsch-Gordan 
sum over the internal quantum number $\nu$ for the color singlet
state $R\bar{R}\rangle$ \cite{deSwart,Rebbi} built from
the states $|R\nu\rangle$ and $|\bar{R}\bar{\nu}\rangle$.

The crossing operator $U_{ts}$   from the $s$-channel basis to the 
$t$-channel one 
can be calculated with the help of the above formulas for the $s$-channel
projectors and similar formulas for the
$t$-channel basis (that can be obtained by permuting
$b\leftrightarrow c$).
However, the crossing operation involves also  
the mixed states $|8_A8_S\rangle$ and $|8_S8_A\rangle$. 
It is convenient to use the linear combinations (\ref{eq:260}), and take
the wave functions for the components $|8_A8_S\rangle$ and $|8_S8_A\rangle$  
in the $s$-channel basis in the form 
\beq 
\langle abcd|8_A8_S\rangle=\frac{1}{\sqrt{40}}f_{abk}d_{kcd}\,,\,\,\,\,\,
\langle abcd|8_S8_A\rangle=\frac{1}{\sqrt{40}}d_{abk}f_{kcd}\,.
\label{eq:500}   
\eeq
Similar formulas for the $t$-channel basis are obtained by
interchanging $b\leftrightarrow c$.
A straightforward calculation  gives
\beq
\begin{pmatrix}
|1 1\rangle\vspace{1.53mm} \\
|8_A8_A\rangle\vspace{1.53mm}\\
|8_S8_S\rangle\vspace{1.53mm}\\
|27 27\rangle
\vspace{1.53mm} \\
|10\overline{10}\rangle
\vspace{1.53mm} \\
|\overline{10}10\rangle
\vspace{1.53mm} \\
|(8_A8_S)_+\rangle
\vspace{1.53mm} \\
|(8_A8_S)_-\rangle
\vspace{.53mm} 
\end{pmatrix}_t
=
\begin{pmatrix}
\frac{1}{8} & \frac{1}{\sqrt{8}} & \frac{1}{\sqrt{8}} &
\frac{3\sqrt{3}}{8} & 
\frac{\sqrt{5}}{4\sqrt{2}} &\frac{\sqrt{5}}{4\sqrt{2}} & 0 & 0 \\ 
\frac{1}{\sqrt{8}} & \frac{1}{2} & \frac{1}{2} &
-\frac{\sqrt{3}}{2\sqrt{2}} & 0 &
0 & 0 & 0 \\ 
\frac{1}{\sqrt{8}} & \frac{1}{2} & -\frac{3}{10} &
\frac{3\sqrt{3}}{10\sqrt{2}} & -\frac{1}{\sqrt{5}} &
-\frac{1}{\sqrt{5}} & 0 & 0\\
\frac{3\sqrt{3}}{8} & -\frac{\sqrt{3}}{2\sqrt{2}} 
& \frac{3\sqrt{3}}{10\sqrt{2}} &
\frac{7}{40} & -\frac{\sqrt{3}}{4\sqrt{10}} & -\frac{\sqrt{3}}{4\sqrt{10}} & 
0 & 0\\
\frac{\sqrt{5}}{4\sqrt{2}} & 0 
& -\frac{1}{\sqrt{5}} & -\frac{\sqrt{3}}{4\sqrt{10}} &
\frac{1}{4}  & \frac{1}{4} & -\frac{1}{\sqrt{2}} &  0\\
\frac{\sqrt{5}}{4\sqrt{2}} & 0 
& -\frac{1}{\sqrt{5}} & -\frac{\sqrt{3}}{4\sqrt{10}} &
\frac{1}{4}  & \frac{1}{4} & \frac{1}{\sqrt{2}} &  0\\
0 & 0 & 0 & 0 & -\frac{1}{\sqrt{2}} &  \frac{1}{\sqrt{2}} & 0 & 0 \\
0 & 0 & 0 & 0 & 0 & 0 & 0 & -1 
\end{pmatrix}
\times
\begin{pmatrix}
|1 1\rangle\vspace{1.53mm} \\
|8_A8_A\rangle\vspace{1.53mm}\\
|8_S8_S\rangle\vspace{1.53mm}\\
|27 27\rangle
\vspace{1.53mm} \\
|10\overline{10}\rangle
\vspace{1.53mm} \\
|\overline{10}10\rangle
\vspace{1.53mm} \\
|(8_A8_S)_+\rangle
\vspace{1.53mm} \\
|(8_A8_S)_{-}\rangle
\vspace{.53mm} 
\end{pmatrix}_s
\label{eq:510}   
\eeq
As one can see the crossing matrix is real and symmetrical.

A straightforward calculation of the $6\times 6$ diffraction matrix in 
the $s$-channel 
basis with the help of the above formulas for the projectors gives
(the order of states is the same as in (\ref{eq:100}))
\beq
\hat{\sigma}(\rho)=
\sigma_{8}(\rho)\times
\begin{pmatrix}
2 & -\frac{1}{\sqrt{2}} & 0  & 0 & 0 & 0\\
-\frac{1}{\sqrt{2}} & \frac{3}{2} & -\frac{1}{2} &
-\frac{1}{\sqrt{6}} & 0 & 0 \\ 
0 & -\frac{1}{2} & \frac{3}{2} & 0 & -\frac{1}{\sqrt{5}} &
-\frac{1}{\sqrt{5}} \\
0 & -\frac{1}{\sqrt{6}} & 0 & \frac{2}{3} &
-\frac{2}{\sqrt{30}} & -\frac{2}{\sqrt{30}} \\
0 & 0 
& -\frac{1}{\sqrt{5}} & -\frac{2}{\sqrt{30}} & 1 & 0 \\
0 & 0 
& -\frac{1}{\sqrt{5}} & -\frac{2}{\sqrt{30}} & 0 & 1
\end{pmatrix}\,,
\label{eq:520}   
\eeq
where $\sigma_{8}(\rho)$ is the dipole cross section for a color
singlet two-gluon system of the size $\rho$.

\end{document}